\documentclass{aastex63}

\submitjournal{AJ}

\shorttitle{Jupiter in the ultraviolet}
\shortauthors{Melin et al.}

\begin{document}

\title{Jupiter in the ultraviolet: acetylene and ethane abundances in the stratosphere of Jupiter from Cassini observations between 0.15 and 0.19 $\mu$m}

\correspondingauthor{Henrik Melin}
\email{henrik.melin@leicester.ac.uk}

\author[ 0000-0001-5971-2633]{Henrik Melin}
\affiliation{School of Physics \& Astronomy \\
University of Leicester \\
University Road \\
Leicester, LE1 7RH, UK}

\author{L. N. Fletcher}
\affiliation{School of Physics \& Astronomy \\
University of Leicester \\
University Road \\
Leicester, LE1 7RH, UK}

\author{P. G. J. Irwin}
\affiliation{University of Oxford, UK}

\author{S. G. Edgington }
\affiliation{Jet Propulsion Laboratory, Pasadena, USA}



\begin{abstract}

At wavelengths between 0.15 and 0.19 $\mu$m, the far-ultraviolet spectrum of Jupiter is dominated by the scattered solar spectrum, attenuated by molecular absorptions primarily by acetylene and ethane, and to a lesser extent ammonia and phosphine. We describe the development of our radiative transfer code that enables the retrieval of abundances of these molecular species from ultraviolet reflectance spectra. As a proof-of-concept we present an analysis of Cassini Ultraviolet Imaging Spectrograph (UVIS) observations of the disk of Jupiter during the 2000/2001 flyby. The ultraviolet-retrieved acetylene abundances in the upper stratosphere are lower than those predicted by models based solely on infrared thermal emission from the mid-stratosphere observed by the Composite Infrared Spectrometer (CIRS), requiring an adjustment to the vertical profiles above 1 mbar. We produce a vertical acetylene abundance profile that is compatible with both CIRS and UVIS, with reduced abundances at pressures $<$1 mbar: the 0.1 mbar abundances are $1.21 \pm 0.07$ ppm for acetylene and $20.8 \pm 5.1$ ppm for ethane. Finally, we perform a sensitivity study for the JUICE UVS instrument, which has extended wavelength coverage out to 0.21 $\mu$m, enabling the retrieval of ammonia and phosphine abundances, in addition to acetylene and ethane. 

\end{abstract}



\section{Introduction} \label{sec:intro}

\noindent  Ultraviolet photons are unable to penetrate the atmosphere of the Earth, and so we need instruments mounted on spacecraft to observe these photons both reflected and emitted by the giant planets. There have been a number of dedicated ultraviolet space telescopes, such as the International Ultraviolet Explorer (IUE), the Hubble Space Telescope (HST), and the Astro 1 \& 2 Space Shuttle missions. In addition, ultraviolet spectrographs often form part of larger instrument suites on spacecraft either performing flybys or entering orbit about a giant planet, e.g., the Voyager 1 \& 2 tour of the outer solar system, the Cassini mission to Saturn, and the Juno mission currently in orbit about Jupiter. 

The ultraviolet reflectance spectrum of Jupiter, observed long-ward of $\sim$0.15 $\mu$m, is dominated by Rayleigh-Raman scattered sunlight from the species contained within the atmosphere, mainly H$_2$, He, CH$_4$, and NH$_3$. This reflectance spectrum is attenuated by wavelength-dependent absorptions by molecules with the amount of absorption being governed by the ultraviolet absorption cross-section, multiplied by the volumetric abundance at a particular altitude. Therefore, if we can determine the amount of absorption present in a spectrum, i.e., the departures from pure Rayleigh scattering at a particular wavelength, we can retrieve the abundances the absorbing species. Additionally, aerosols will also scatter light away from and into the line-of-sight which could potentially significantly reduce the observed radiance, dependent on their properties. 

Acetylene and ethane are both products of methane photochemistry in the stratosphere driven by solar irradiation, yet they display very different meridional profiles \citep[e.g.][]{2007Icar..188...47N}: acetylene peaks at the equator, whilst ethane peaks at high-to-mid latitudes. This has been attributed to the different life-times of the two species, with acetylene having life-times of hundreds of days, and therefore responds to short-term atmospheric dynamics, compared to several years for ethane. \cite{2018Icar..305..301M} characterised the long-term behaviour of acetylene and ethane over a period of a half jovian year using emission features in the mid-infrared, as measured by the TEXES spectrograph \citep{2002PASP..114..153L}, and showed that whilst ethane does not change significantly, consistent with a very long life-time, acetylene showed significant variability, with the global distribution changing from asymmetric about the equator in 2012 to symmetric in 2017. This shows that acetylene is sensitive to short-term dynamics, such as vertical transport possibly driven by changes in convection in the deeper troposphere.


Acetylene was first discovered in the ultraviolet at Jupiter by \cite{1980ApJ...236L..39O}, using observations from the IUE observatory in 1979, noting the resemblance between broad features in the observed spectrum around 0.17 $\mu$m, and the acetylene absorption cross-section measured in the laboratory. Table \ref{previous} summarises the abundances of acetylene and ethane determined by previous studies in the ultraviolet.


The IUE had a spectral coverage between 0.145 to 0.315 $\mu$m, and so encapsulated absorptions by a wide range of species. \cite{1985Icar...63..222W} examined the retrievability of a whole host of methane photochemical products from the ultraviolet Jupiter spectra, including species like methylacetylene (C$_3$H$_4$) and hydrogen sulphide (H$_2$S), but concluded that only acetylene, ethane, and ammonia could be retrieved with relative confidence. Since this study, improved ammonia ultraviolet absorption cross-section laboratory measurements \citep[e.g.][]{1998PSS...47..261C, 2007ApJ...657L.117L} have enabled the retrieval of ammonia. Both ammonia and phosphine are tracers of vertical bulk motions in the atmosphere \citep{2009Icar..202..543F, 2017GeoRL..44.5317L}. These species reside in the deep troposphere, but vertical transport, driven by atmospheric convection, move them above the cloud decks where they can be observed, and subsequently re-distributed by atmospheric dynamics. 

\cite{1998Icar..133..192E, 1999Icar..142..342E} analysed ultraviolet spectra obtained using the Faint Object Spectrograph (FOS) mounted on the HST, with wavelength coverage between 0.16 and 0.23 $\mu$m, determining abundances for acetylene and ammonia, at around 80 mbar and 120 mbar, respectively (see Table \ref{previous}). Using a photochemical model, they calculated the ammonia eddy diffusion coefficient, a measure of the strength of the vertical mixing. They found that the ammonia abundances varied with latitude, suggesting the vertical mixing is dependent on latitude. At wavelengths longer than 0.21 $\mu$m, they noted that the inclusion of Raman scattering and an unknown continuum absorber are required to explain a set of features at these wavelengths. 

A subset of the HST FOS data-set was re-visited by \cite{2003Icar..163..414B}. Their modelling required an ethane abundance of 150 ppb at 120 to 300 mbar in the upper troposphere, disagreeing with previous measurements in the ultraviolet (see Table \ref{previous}) and disagreeing with observations in the mid-infrared. This issue may in part be driven by the FOS {\it red leak}, that increases noise at wavelength were ethane is absorbing. They concluded that near-simultaneous observations in both the ultraviolet and the mid-infrared may be required to investigate whether the two wavelength regions produce truly compatible results. It is a challenge that motivates the present study. 

The ultraviolet wavelength range can also be used to measure the atmospheric attenuation of either sunlight or starlight during occultation \citep[e.g. see][]{1990RvGeo..28..117S}. \cite{2010Icar..208..293G} analysed ultraviolet stellar occultation observations of Jupiter obtained during the New Horizon flyby in 2007, deriving abundances of methane, acetylene, and ethane at pressure levels of 1-10 $\mu$bar. They derived a different altitude of the methane homopause, compared to the Voyager 2 results of \cite{1981JGR....86.5715F}, suggesting long-term changes in the vertical mixing.

The Astro-1 and Astro-2 missions flown onboard the Space Shuttle ({\it Columbia} in 1990 and {\it Endeavour} in 1995, respectively), carried the Hopkins Ultraviolet Telescope (HUT), an instrument with wavelength coverage between 0.145 and 0.185 $\mu$m. From observations of the equatorial region of Jupiter, \cite{1995ApJ...454L..65M} retrieved abundances of acetylene and ethane in the stratosphere. They included diacetylene (C$_4$H$_2$) as a free parameter in their fitting procedure, but were unable to unambiguously detect its presence in the Astro HUT spectra.


Here, we describe the development of our radiative transfer code to be able to model and retrieve atmospheric abundances from ultraviolet reflectance spectra. We then apply this model to observations of Jupiter by the Cassini Ultraviolet Imaging Spectrograph (UVIS), comparing them to results derived from mid-infrared observations acquired during the same time period. Finally we discuss how this model can be applied to data taken by current and future spacecraft missions.

\begin{table}[]
\begin{tabular}{llll}
\hline
Reference & Observatory & C$_2$H$_2$ (ppb)  & C$_2$H$_6$ (ppm)   \\
\hline
\cite{1980ApJ...236L..39O}	& IUE   & 22 & \\
\hline
\cite{1983ApJ...266..415G} 	& IUE &  $100\pm 10$ & 6.6$\pm$5.3 \\
\hline
\cite{1985Icar...63..222W}		& IUE &  $30 \pm 10$ & 2 \\
\hline
\cite{1990JGR....9510365M} & IUE & 33.5  & \\
\hline
\cite{1995ApJ...454L..65M} 	& Astro-1   & 39$\pm$3  & 3$\pm$1      \\
					 	& Astro-2 &  28$\pm$3  & 3$\pm$1      \\

\hline
\cite{1998Icar..133..192E} 	& HST   & 10-30  & \\

\cite{2003Icar..163..414B}	& HST	& 40 & \\

\hline
\end{tabular}
\caption{A summary of previous acetylene and ethane abundances derived from ultraviolet observations. These values are estimated to be at $\sim$80 mbar, but may actually be at higher altitudes - see Section \ref{secdiss}. \label{previous}}
\end{table}



\begin{figure}
\centering
\includegraphics[width=0.8\textwidth]{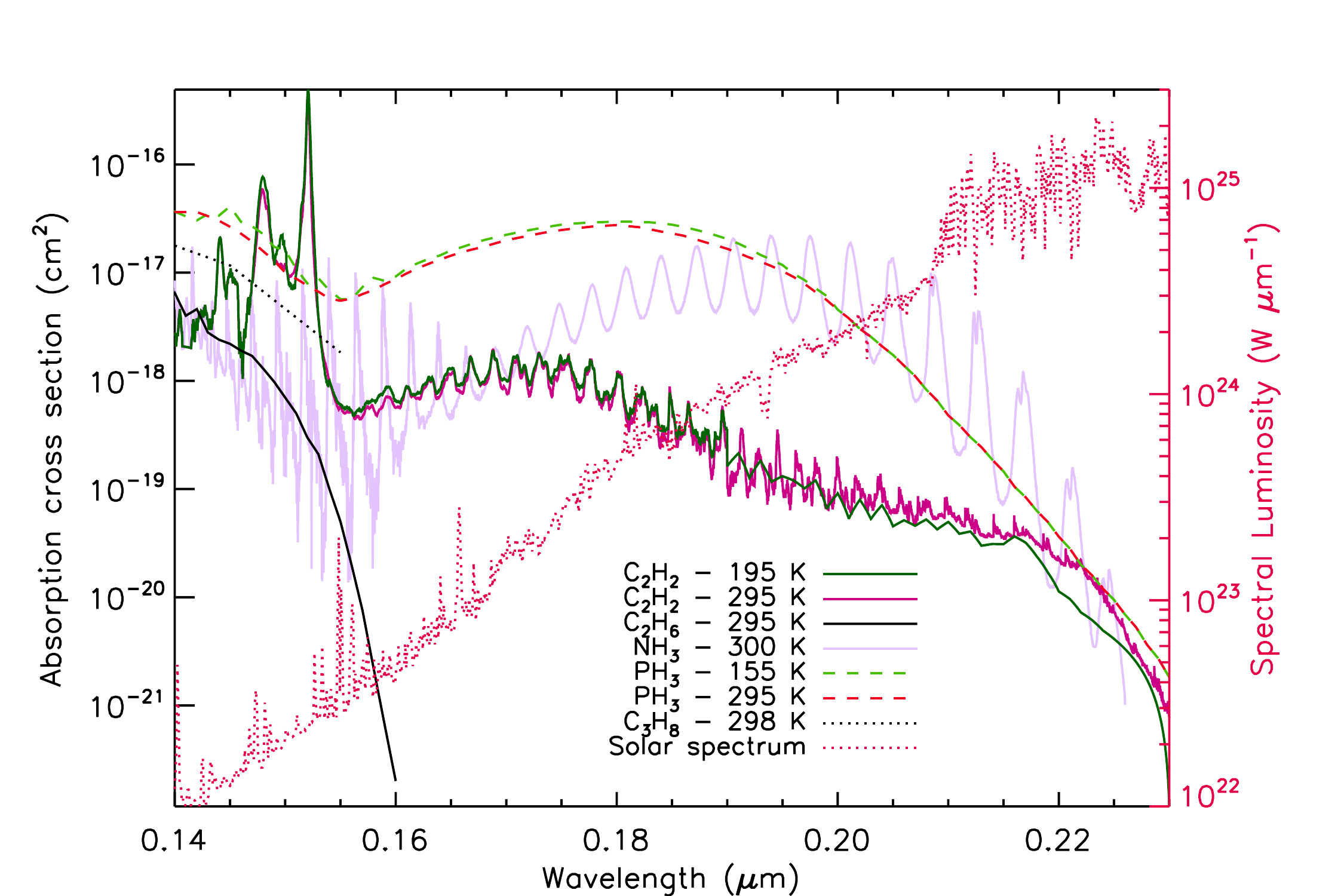}
\caption{The ultraviolet absorptions cross-sections relevant to this study: acetylene, ethane, ammonia, and phosphine, with the respective references listed in Table \ref{xtable}. When available, the cross-sections at different temperatures are shown. The spectral luminosity of the high spectral resolution \cite{2009GeoRL..36.1101W} solar reference spectrum is shown as the red dotted line (using the right-hand axis). \label{xsect}}
\end{figure}

\begin{table}[]
\begin{tabular}{llll}
\hline 
Molecule & Reference  & $T$ (K) & Coverage ($\mu$m)  \\
\hline 
\hline
C$_2$H$_2$ (acetylene)  	& \cite{1991JGR....9617529S}  	& 195  & 0.137 $\rightarrow$ 0.190  \\
C$_2$H$_2$  	& \cite{1991JGR....9617529S}  	& 295 & 0.137 $\rightarrow$ 0.190  \\
C$_2$H$_2$  	& \cite{2000PSS...48..463B}  	& 195 & 0.190 $\rightarrow$ 0.225 \\
C$_2$H$_2$  	& \cite{2000PSS...48..463B}  	& 295 & 0.190 $\rightarrow$ 0.225 \\
\hline
C$_2$H$_6$  (ethane)     & \cite{2001ApJ...551L..93L} 	& 295 & 0.120 $\rightarrow$ 0.160 \\
\hline
NH$_3$ (ammonia)		& \cite{2007ApJ...657L.117L} 	& 300 & 0.140 $\rightarrow$ 0.226 \\
\hline
PH$_3$ (phosphine)		& \cite{1991JGR....9617519C} 	& 155 & 0.127 $\rightarrow$ 0.200  \\
PH$_3$ 		& \cite{1991JGR....9617519C} 	& 295 & 0.120 $\rightarrow$ 0.230 \\
\hline
H$_2$O (water)	& \cite{jpl18} & 298 &  0.120 $\rightarrow$ 0.200 \\
\hline
CH$_4$ (methane)			& \cite{1991JGR....9617519C} & 295 & 0.120 $\rightarrow$ 0.142 \\
\hline
CO (carbon monoxide)		& \cite{1993CP....170..123C} & 298 & 0.010 $\rightarrow$ 0.180 \\
\hline
C$_4$H$_2$ (diacetylene) & \cite{2009PSS...57...10F} & 295 & 0.110 $\rightarrow$ 229 \\
\hline
GeH$_4$ (germane) & \cite{1991JGR....9617519C} & 295 & 0.114 $\rightarrow$ 0.180 \\
\hline
C$_3$H$_8$ (propane) & \cite{1993CP....173..209A} & 298 & 0.05 $\rightarrow$ 0.155 \\
\hline

\end{tabular}
\caption{The sources of the measured ultraviolet absorption cross-sections included in the updated NEMESIS radiative transfer and retrieval code, listing the molecular species, source reference, temperature ($T$) and the wavelength coverage.}
\label{xtable}
\end{table}

\section{NEMESIS Development \& Modelling}

We employ the NEMESIS radiative transfer and atmospheric retrieval code \citep{2008JQSRT.109.1136I} to model the ultraviolet reflectance spectra of Jupiter. This code has two basic operational modes: a) {\it the forward model} -- given a set of atmospheric abundance and temperature profiles it will calculate the resulting spectrum at the top of the atmosphere, and b) {\it the retrieval} -- given an observed spectrum, NEMESIS will calculate the atmospheric profiles that are consistent with the observations, using optimal estimation to iterate towards the solution \citep{2000imas.book.....R}. 

NEMESIS calculates multiple scattering via a matrix operator approach \citep{1973ApOpt..12..314P}, also known as doubling-and-adding, alongside Rayleigh scattering from the jovian gases. We used five zenith angles and $N$ Fourier components to cover the azimuth variation, where $N$ is set adaptively from the viewing zenith angle, $\theta$, as $N = int(\theta/3)$. NEMESIS employs the plane-parallel approximation whilst using the solar and viewing zenith angles and azimuth angles calculated for the observed point on the planet \citep[e.g.][]{1991P&SS...39..671D}.

In the absence of molecular absorption and scattering by aerosols, the ultraviolet spectrum will display a solar reflection, governed by Rayleigh scattering. However, there are a number of constituents in the atmosphere that absorb ultraviolet light, and the abundances and the ultraviolet absorption cross-section of these species determine the amount of absorption. Additionally, aerosols in the atmosphere of Jupiter will absorb and scatter light from the observer's line-of-sight, acting as to reduce the observed radiance. Because this region is strongly dominated by Rayleigh scattering of the solar spectrum, there are no molecular emissions included in these calculations, and so neither the correlated-$k$ approximation or line-by-line calculations are used.


This study required development of the core NEMESIS code in the following ways: adding a high resolution ultraviolet solar spectrum, adding ultraviolet absorption cross-sections, and a means by which to toggle the absorbing species in our retrieval code. These developments are outlined in greater details below. 

\subsection{Solar spectrum}

At infrared wavelengths the \cite{2010JQSRT.111.1289C} solar reference spectrum provides the spectral resolution required for accurate modelling of reflectance spectra. However, at ultraviolet wavelengths this spectrum is much too coarse for accurate Rayleigh scattering calculations. We added the \cite{2009GeoRL..36.1101W} reference spectrum, shown as the red dotted line in Figure \ref{xsect}, providing wavelength coverage between 0.0001 $\mu$m and 0.240 $\mu$m at a spectral resolution of 0.1 nm, which is  higher than the spectral resolution of the data-set used in this study (described in more detail in Section \ref{obssec}) and so is appropriate for use here. NEMESIS convolves the solar spectrum to the resolution of the observation in questions and computes the Rayleigh scattering using the formulation of Sromovsky ({\it private communication}) considering scattering from H$_2$, He, CH$_4$ and NH$_3$. 

\subsection{Ultraviolet absorption cross-section \label{xsectsect}}

The Rayleigh-scattered sunlight is attenuated by absorption by the molecular species present in the atmosphere and by scattering by aerosols. The amount of absorption by each species is a function of abundance and the wavelength-dependent ultraviolet absorption cross-section. The larger the cross-section, the larger the absorption of light, and the lower the observed radiance becomes. The cross-sections that were added to the NEMESIS code are listed in Table \ref{xtable}, and the ones relevant for the current study are plotted as a function of wavelength in Figure \ref{xsect}. The cross-sections were sourced from the database of  \cite{2013ESSD....5..365K}\footnote{\url{http://satellite.mpic.de/spectral_atlas}}.

The temperature of the troposphere and stratosphere of Jupiter ranges between 100 and 200 K (e.g. see Figure \ref{apriori}). The temperature range of the cross-section measurements in Table \ref{xtable} and Figure \ref{xsect} are measurements between 155 and 300 K, effectively rendering them to be single temperature measurements when applied to the atmosphere of Jupiter. Both acetylene and phosphine have two cross-section temperature measurements, and the difference between the two temperatures is very slight for each of these species (see Figure 1). If a temperature at a particular pressure level is in-between the two temperatures that the cross-sections were measured at, we interpolate between the two values. The lack of low temperature ultraviolet cross-sections highlights that there is plenty of scope for further experimental work. 

\begin{figure}
\centering
\includegraphics[width=0.8\textwidth]{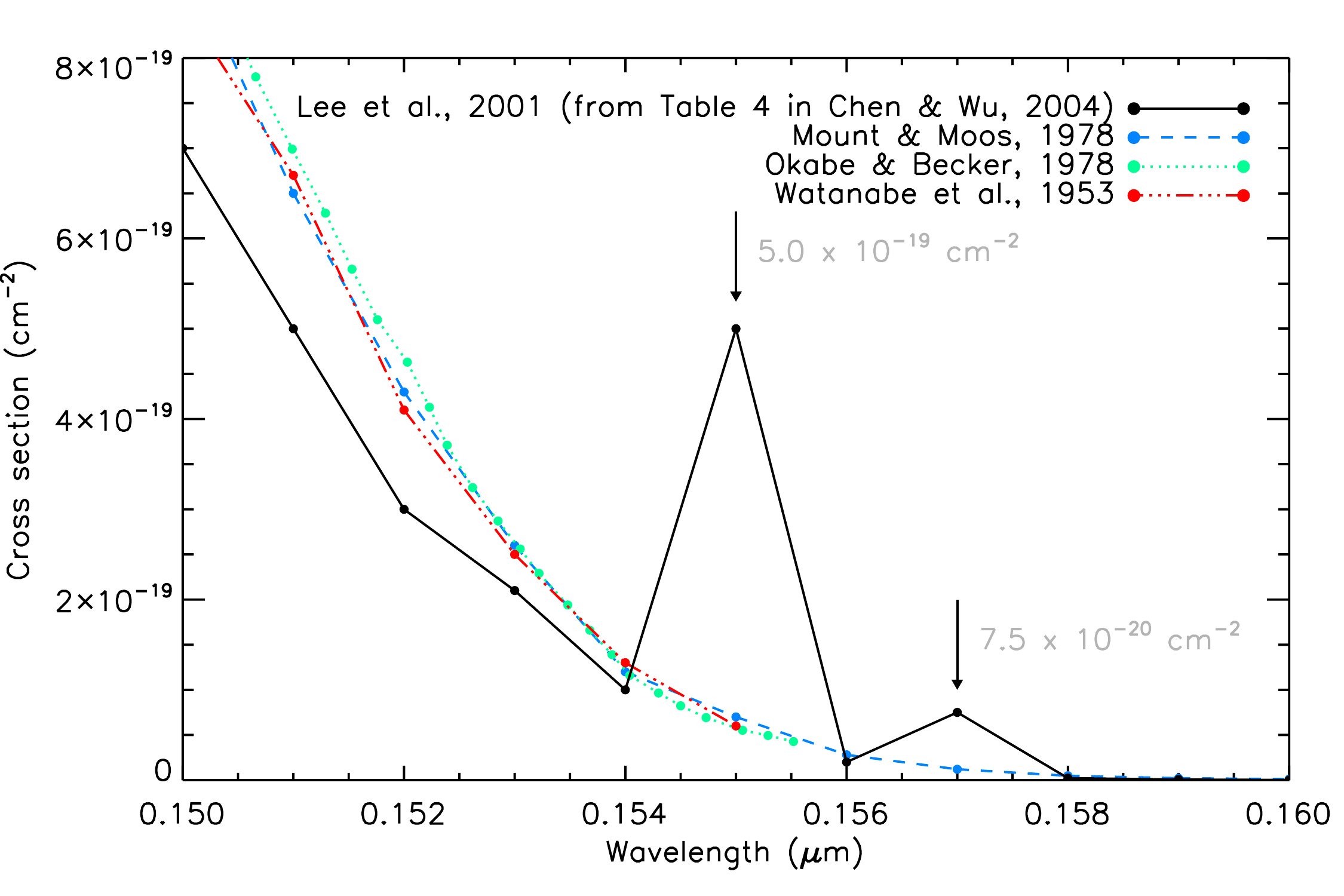}
\caption{The ultraviolet absorption cross-section of ethane between 0.15 and 0.16 $\mu$m from four different experiments. The arrows indicate the location of the anomalous data points listed in   \cite{2004JQSRT..85..195C}, which purports to list the results of \cite{2001ApJ...551L..93L}. These are both inconsistent with the plots in \cite{2001ApJ...551L..93L} and with previous measurements. The correct values for the ethane cross-section are listed in Table \ref{ethanexsect}. \label{ethanexsectf}}
\end{figure}

\begin{table}[]
\begin{tabular}{ll||ll}
\hline
Wavelength (nm) & Cross section (cm$^{-2}$) & Wavelength (nm) & Cross section (cm$^{-2}$)    \\
\hline
      120 & 2.3 $	\times 10^{-17} $ &       140 & 6.5 $	\times 10^{-18} $ \\ 
      121 & 2.2 $	\times 10^{-17} $ &       141 & 4.0 $	\times 10^{-18} $ \\ 
      122 & 2.3 $	\times 10^{-17} $ &       142 & 4.6 $	\times 10^{-18} $ \\ 
      123 & 2.1 $	\times 10^{-17} $ &       143 & 2.8 $	\times 10^{-18} $ \\ 
      124 & 2.3 $	\times 10^{-17} $ &       144 & 2.4 $	\times 10^{-18} $ \\ 
      125 & 2.4 $	\times 10^{-17} $ &       145 & 2.2 $	\times 10^{-18} $ \\ 
      126 & 2.6 $	\times 10^{-17} $ &       146 & 1.9 $	\times 10^{-18} $ \\ 
      127 & 2.5 $	\times 10^{-17} $ &       147 & 1.7 $	\times 10^{-18} $ \\ 
      128 & 1.9 $	\times 10^{-17} $ &       148 & 1.3 $	\times 10^{-18} $ \\ 
      129 & 1.8 $	\times 10^{-17} $ &       149 & 9.8 $	\times 10^{-19} $ \\
      130 & 2.7 $	\times 10^{-17} $ &         150  &    7.0$	\times 10^{-19} $ \\ 
      131 & 2.5 $	\times 10^{-17} $ &         151   &  5.0$	\times 10^{-19} $ \\ 
      132 & 2.8 $	\times 10^{-17} $ &         152   &  3.0$	\times 10^{-19} $ \\ 
      133 & 2.3 $	\times 10^{-17} $ &         153   & 2.1$	\times 10^{-19} $ \\ 
      134 & 2.3 $	\times 10^{-17} $ &         154  &  1.0$	\times 10^{-19} $ \\ 
      135 & 1.8 $	\times 10^{-17} $ &         155 &   5.0$	\times 10^{-20} $ \\ 
      136 & 1.9 $	\times 10^{-17} $ &         156  &  2.0$	\times 10^{-20} $ \\ 
      137 & 1.2 $	\times 10^{-17} $ &         157  &  7.5$	\times 10^{-21} $ \\ 
      138 & 1.2 $	\times 10^{-17} $ &         158  &  2.2$	\times 10^{-21} $ \\ 
      139 & 6.0 $	\times 10^{-18} $ &         159  &  6.6$	\times 10^{-22} $ \\ 
 & &              160  &  2.0$	\times 10^{-22} $ \\ 
       \hline
\end{tabular}
\caption{The corrected tabulation of \cite{2001ApJ...551L..93L} (Robert Wu, private communication). }
\label{ethanexsect}
\end{table}

The cross-section of ethane merits a closer examination. The most recent measurements beyond 0.15 $\mu$m are those of \cite{2001ApJ...551L..93L}. However, these authors do not tabulate these values, opting to only to plot them (bold line in Figure 1 in their paper). The values of \cite{2001ApJ...551L..93L} are tabulated in Table 4 in \cite{2004JQSRT..85..195C}. However, there are two anomalous spectral features listed in \cite{2004JQSRT..85..195C} that are not present in the original \cite{2001ApJ...551L..93L} study, one peak at 0.155 $\mu$m and one at 0.157 $\mu$m. This is shown in Figure \ref{ethanexsectf}, along with the measured cross-section from \cite{1978ApJ...224L..35M}, \cite{1963JChPh..39.2549O}, and \cite{watc2h2}. The two peaks indicated by the arrows are unlikely to be physical since \cite{2001ApJ...551L..93L} states that the ethane cross-sections they measure are nearly identical to those of \cite{1978ApJ...224L..35M}, which is also absent of the features at 0.155 and 0.157 $\mu$m, and they disagree wildly with all previous measurements. It turns out that these two cross-sections have erroneous exponents making them too large by a factor of ten (Robert Wu, private communication). The correct values are listed in Table \ref{ethanexsect}.





\begin{figure}
\centering
\includegraphics[width=0.8\textwidth]{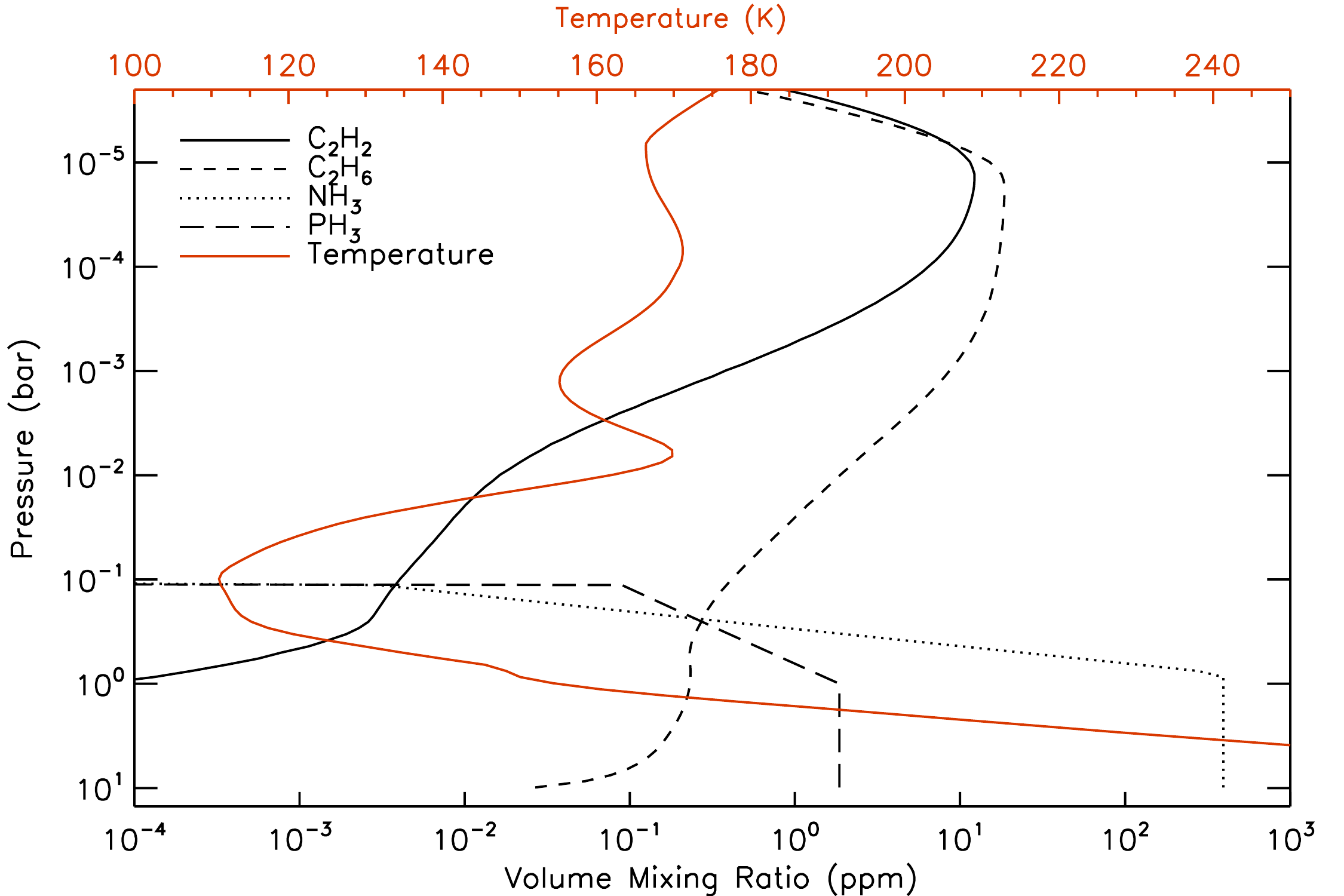}
\caption{The equatorial a-priori vertical profiles of acetylene, ethane, ammonia, phosphine and temperature used in this study. These were obtained from observations by the Cassini CIRS instrument during the Jupiter flyby \citep{2010P&SS...58.1667N, 2016Icar..278..128F}. The Quasi Quadrennial Oscillation (QQO) is clearly visible in the stratospheric temperature profile at pressures $<$10 mbar. \label{apriori}}
\end{figure}


\subsection{Aerosols}

For ultraviolet observations of Jupiter's stratosphere, aerosols have been shown to play an important role \citep[e.g.][]{1972ApJ...173..451A}, and can significantly alter the shape of the observed spectrum. Here, we use the aerosol properties derived by \cite{1998Icar..133..192E}, consistent with their HST FOS observations: a particle size of 0.3 $\mu$m distributed with a variance of 0.05 $\mu$m, a refractive index of $1.4 + 0.3i$, and an optical depth of unity at 1 bar. This particle sizes are also consistent with \cite{1986Icar...65..218T} who analysed observations of Jupiter obtained long-ward of 0.22 $\mu$m from the IUE. The base of the vertical distribution is at 10 mbar, with a fractional scale height of 1. 

The extinction cross-section and single-scattering albedo (as a function of wavelength) of the aerosols was calculated using Mie theory \citep[e.g.][]{1974SSRv...16..527H}, and from these optical properties and vertical distribution we calculate the phase behaviour for the aerosols, represented by double Henyey-Greenstein scattering functions. 

In Section \ref{aersec} we explore the aerosol properties to which Jupiter's reflectance spectrum is sensitive. 



\subsection{Raman scattering}

In addition to Rayleigh scattering, solar photons can also undergo Raman scattering, in which a photon exchanges energy with a molecule, emerging with a wavelength either longer or shorter than the original one. The shift in wavelength is governed by the rotational and vibrational states available in the scattering molecule: mainly molecular hydrogen in the atmospheres of the giant planets \citep[e.g.][]{2005Icar..173..254S}.

\cite{1999Icar..142..342E} and \cite{1999Icar..142..324B} showed that there are strong spectral features long-ward of 0.21 $\mu$m present in HST FOS spectra that are attributable to Raman scattering. At wavelengths shorter than 0.21 $\mu$m, Raman scattering contributed only 2-3\% of the total signal, whereas long-ward of 0.21 $\mu$m it contributes 3-6\% of the overall signal. The Cassini UVIS wavelength coverage ends at $\sim$0.19 $\mu$m and has an uncertainty in the radiance significantly larger than the contribution of Raman scattering calculated by \cite{1999Icar..142..324B}. Therefore we do not include Raman scattering in the present study. However, in scenarios where the uncertainty of the data is considerably lower and when the wavelength coverage extends beyond 0.21 $\mu$m, Raman scattering needs to be included to achieve accurate and physical retrievals

\subsection{Reference atmosphere}

The Cassini Composite Infrared Spectrometer \cite[CIRS,][]{2004SSRv..115..169F} provided near-simultaneous observations of Jupiter's atmosphere to those of Cassini UVIS. From these, vertical profiles of temperature and abundances of, amongst others, acetylene, ethane, ammonia, and phosphine have been retrieved \citep{2010P&SS...58.1667N, 2016Icar..278..128F}. The vertical abundance and temperature profiles derived by CIRS at the equator of Jupiter are shown in Figure \ref{apriori}. Acetylene and ethane both peak at about 10 $\mu$bar, whilst both ammonia and phosphine peak at pressures $>$1 bar and are completely absent by 100 mbar due to condensation and photochemsitry. The Cassini CIRS instrument peaks in sensitivity to acetylene at about 10 mbar, and for ethane at about 5 mbar \citep{2010P&SS...58.1667N}. 

The infrared and ultraviolet Cassini observations of Jupiter were obtained at about the same time, and so we can use the CIRS profiles to generate NEMESIS forward models in the ultraviolet wavelength range, to make a prediction of what we may expect an ultraviolet spectrum to appear like, and it enables us to rule out temporal effects between the two data-sets. This is described in the next section.


\begin{figure}
\centering
\includegraphics[width=0.95\textwidth]{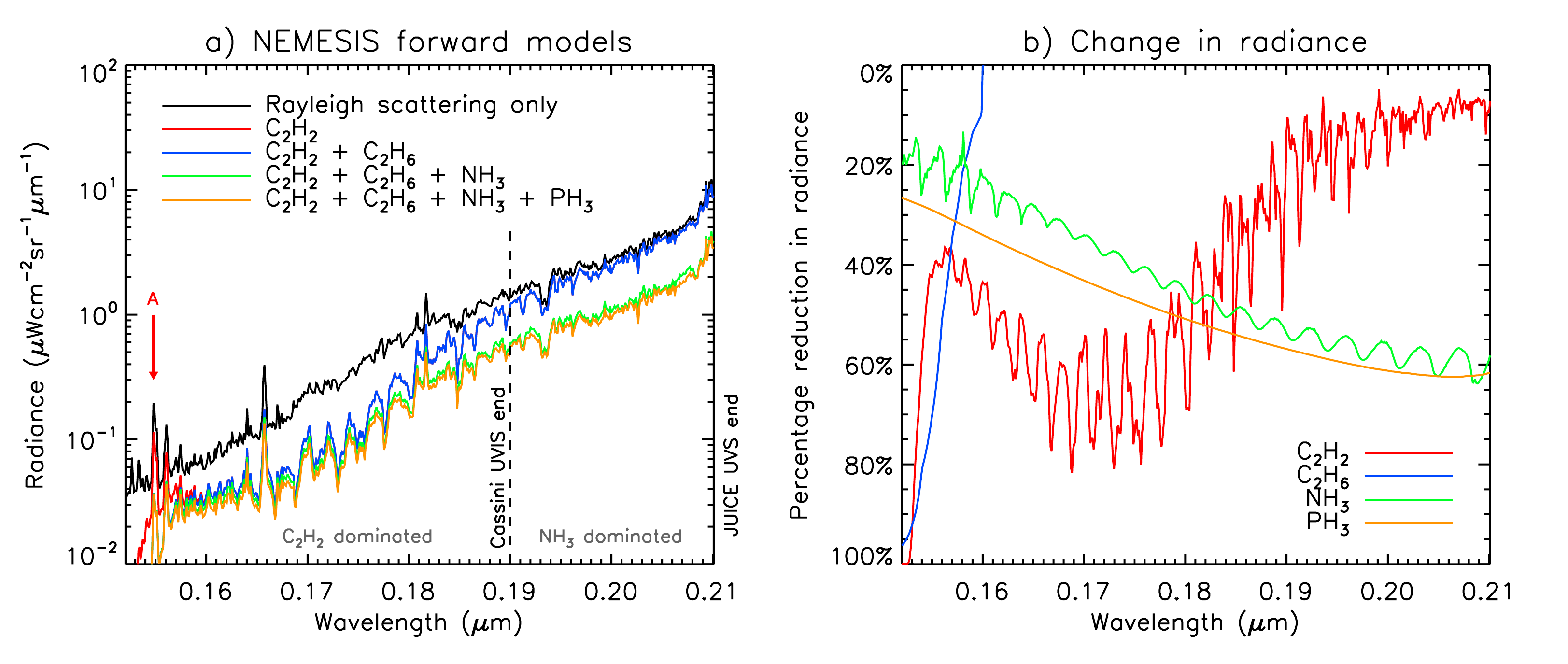}
\caption{a) Five NEMESIS forward models of Jupiter, between 0.15 and 0.21 $\mu$m, showing how the absorption features of acetylene, ethane, ammonia, and phosphine all add as to reduce and attenuate the Rayleigh-scattered sunlight. The vertical dashed line indicates the end of the wavelength coverage of Cassini UVIS (0.19 $\mu$m), whilst JUICE UVS has coverage out to 0.21 $\mu$m. The red arrow (A) indicates the distinct ethane absorption feature at 0.155 $\mu$m. Note that the blue and red line coincide with each other for most of the spectral range. b) The change in radiance produced by modelling the absorption of individual species compared to the Rayleigh scattering only case. The change in radiance broadly reproduces the shape of the absorption cross sections in Figure \ref{xsect}, as expected.  \label{fmadd}}
\end{figure}

\subsection{Forward models}

Using the equatorial profiles obtained by Cassini CIRS, shown in Figure \ref{apriori}, we can generate NEMESIS forward models that calculate the observed spectrum emanating from the top of the atmosphere in the FUV wavelength range. Figure \ref{fmadd}a shows the modelled radiance between 0.15 and 0.21 $\mu$m for different combinations of molecular absorbers, from Rayleigh scattering alone, to the inclusion of all the different relevant species: acetylene, ethane, ammonia, and phosphine. It shows that acetylene absorptions dominates between 0.15 and 0.19 $\mu$m (red and blue curves), with distinct repeating features between 0.17 and 0.18 $\mu$m. In the narrow region at $\sim$0.155 $\mu$m ethane creates a sharp and distinct absorption feature (indicated as A). At wavelengths longer than 0.19 $\mu$m the spectrum is dominated by ammonia absorptions, with phosphine having a minor contribution over the entire spectral range (green and orange curves). 

The forward models in Figure \ref{fmadd}a, using vertical profiles from CIRS, provide an expectation of what ought to be observed in actual FUV spectrum from Jupiter. The predicted absorption features are significant, and so we may expect to be able to retrieve abundances of some, or all of these species. In the next sections, we will put this to the test by analysing ultraviolet Cassini observations of Jupiter.  

Figure \ref{fmadd}b shows the percentage reduction in radiance of a forward model with absorption by a single species turned on in NEMESIS compared to the Rayleigh scattering only case. It shows that the radiance is reduced according to the ultraviolet cross section of each species (Figure \ref{xsect}). This indicates that the NEMESIS implementation is working as expected.

\begin{figure}
\centering
\includegraphics[width=0.8\textwidth]{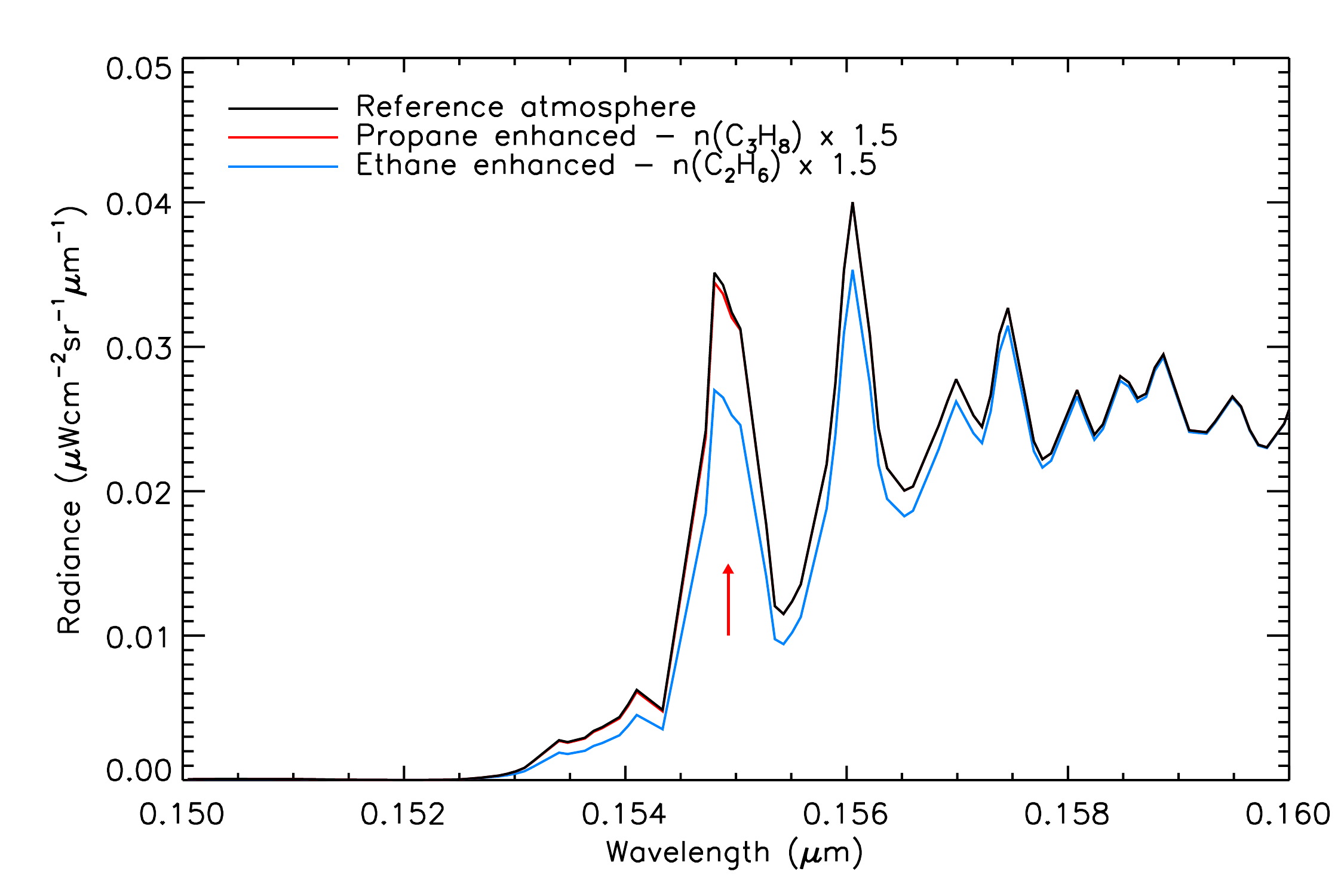}
\caption{The effects of enhancing the abundance of ethane and propane by 50\% in the NEMESIS forward models. Propane has a minimal effect on the observe spectrum, and the feature at 0.155 $\mu$m is dominated by ethane absorptions. \label{propaneinc}}
\end{figure}

The spectral feature seen at 0.155 $\mu$m in Figure \ref{fmadd}a is a signature of ethane. However, propane (C$_3$H$_8$) has a large ultraviolet cross-section with a shape very similar to that of ethane (dotted black line in Figure \ref{xsect}), and so the presence of propane in Jupiter's stratosphere could potentially adversely affect the retrieval of the ethane abundances. We investigate the effect of propane by producing three NEMESIS forward models centred near the 0.155 $\mu$m ethane feature. The first model contains the the abundance of ethane from the reference atmosphere (Figure \ref{apriori}) and the vertical abundance of propane from \cite{2005JGRE..110.8001M}. In the second model, propane is enhanced, multiplied by 1.5 at all altitudes, and in the third model ethane is enhanced by a factor 1.5 at all altitudes. The results of these models are shown in Figure \ref{propaneinc}. It shows that enhancing propane has a minimal effect on the spectrum, reducing it by 2\% at 0.155 $\mu$m. The ethane enhanced atmosphere produces a significant reduction of the observed radiance, reducing the radiance by 30\% at 0.155 $\mu$m. Since the 0.155 $\mu$m feature is clearly dominated by the abundance of ethane, propane is not included in the subsequent analysis.

\begin{figure}
\centering
\includegraphics[width=0.9\textwidth]{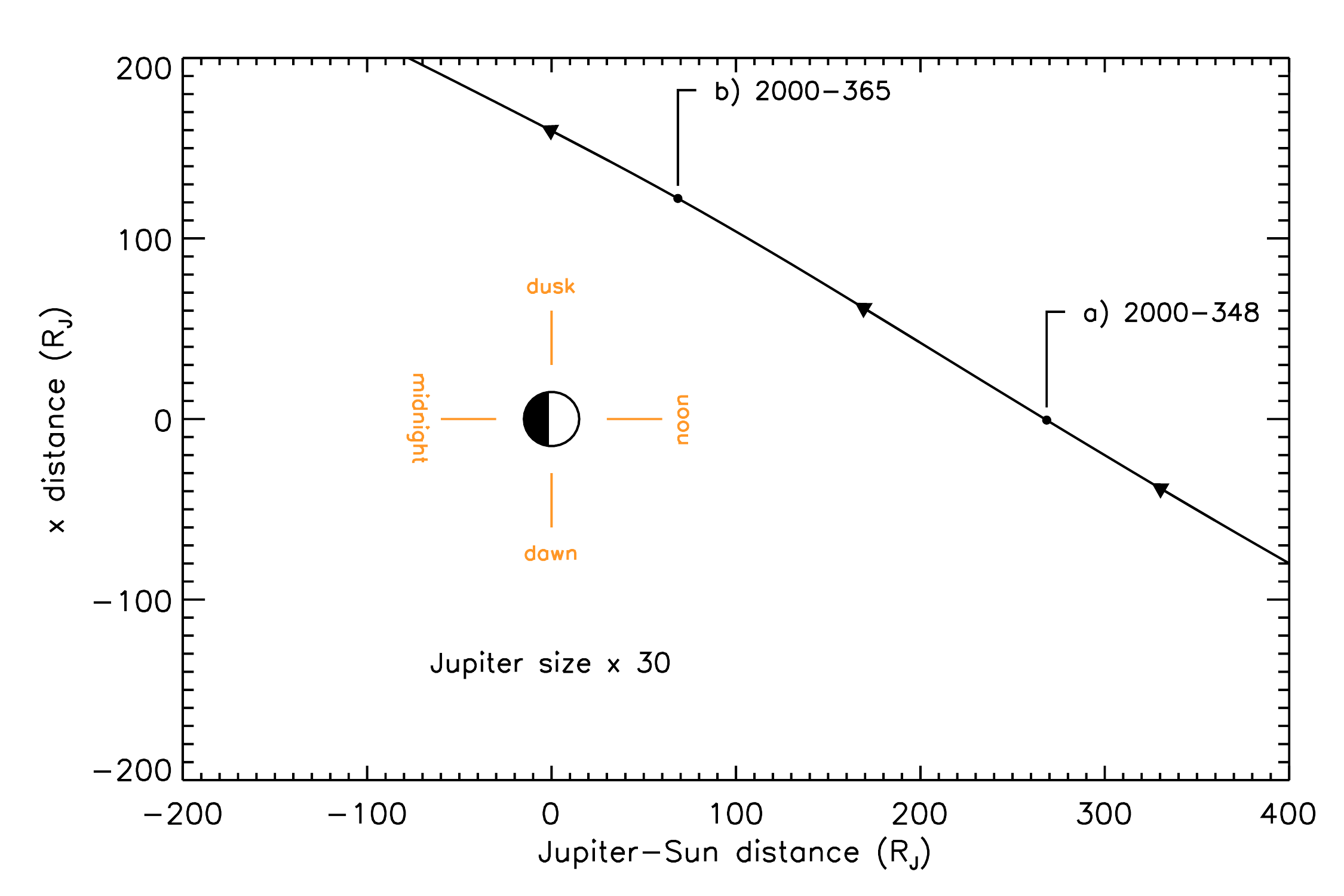}
\caption{The trajectory of Cassini in the Jupiter-Sun frame during the flyby in 2000, as seen from above the north pole of the planet. This study examines two sets of Casini UVIS data: a) data obtained at local noon at very low spatial resolution but at high spectral resolution, and b) the highest spatial resolution observations obtained at closest approach to Jupiter, at reduced spectral resolution. The arrows along the orbit of Cassini are separated by two weeks ($\sim$34 Jupiter rotations). \label{geometry}}
\end{figure}

\begin{figure}
\centering
\includegraphics[width=0.9\textwidth]{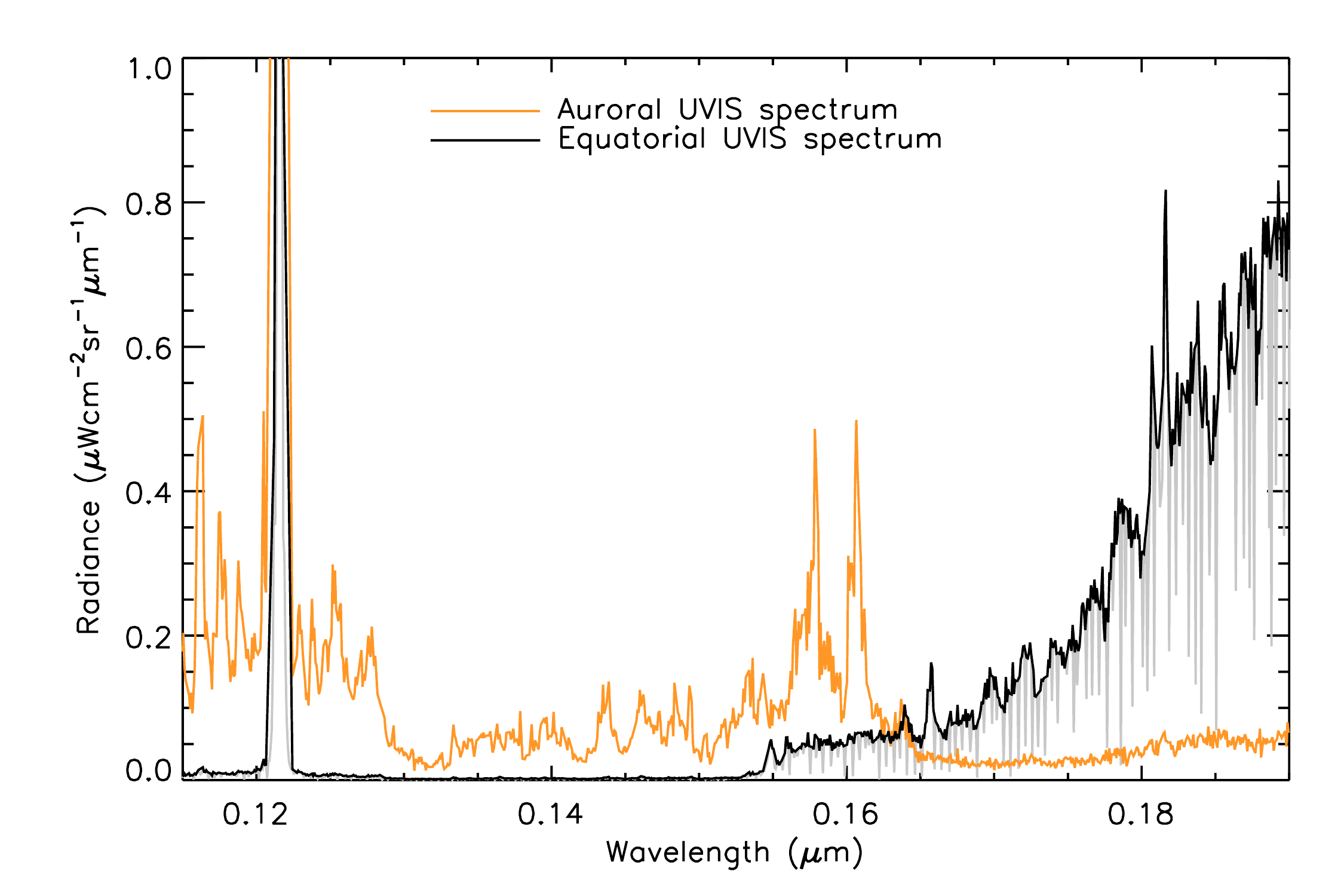}
\caption{The observed equatorial (black) and polar (orange) spectrum of Jupiter observed in the far-ultraviolet by Cassini UVIS. The grey line shows the spectrum, including the bad pixels, which were processed out to produce the final spectrum in black. Note that at high latitudes, the auroral spectrum will add to the solar reflectance spectrum at around 0.16 $\mu$m.  \label{example}}
\end{figure}

\section{Cassini Observations \label{obssec}}

\noindent In late 2000 the Cassini spacecraft approached Jupiter to receive a gravity assist en route to Saturn -- the trajectory in the Jupiter-Sun frame is shown in Figure \ref{geometry}, as seen from above the north pole of the planet. During this interval all the remote sensing instruments on-board Cassini obtained images and spectra of the planet, from the mid-infrared, through the visible, to the extreme ultraviolet. 

The Cassini Ultraviolet Spectrograph \citep[UVIS, ][]{2004SSRv..115..299E} contains two spectrographs, one covering the extreme ultraviolet, the other the far ultraviolet (FUV). The FUV channel has spectral coverage between 0.11 $\mu$m and 0.19 $\mu$m over 1024 spectral pixels, and a spectral resolution of 0.25 nm. The 64 spatial pixels cover 1 $\times$1.5 mrad each, and extended two dimensional spatial coverage is built up by scanning the slit across the object. In this manner a spectral cube can be generated, with two spatial dimensions and one spectral dimension. By applying the Navigation and Ancillary Information Facility (NAIF) spacecraft pointing code \citep{1996P&SS...44...65A}, we calculated the pointing geometry for any particular pixel at any particular time, retrieving parameters such as latitude, longitude, emission angle, solar phase angle, and solar incidence angle.

The Cassini UVIS data is available\footnote{https://atmos.nmsu.edu/pdsd/archive/data/co-j-uvis-2-spec-v12/} on the Planetary Data System (PDS), with the Jupiter encounter covered in archives \verb|couvis_0001| and \verb|couvis_0002|. The sensitivity of the UVIS FUV detector changed over the duration of the Cassini mission, and this is encapsulated in the UVIS calibration routines. We applied the latest version (Version 4) of the calibration software available on the PDS. The software calibrates the data to units of angular surface brightness [kR \AA$^{-1}$], whilst our retrieval code expects units of spectral radiance [W cm$^{-2}$ sr$^{-1}$ $\mu$m$^{-1}$]. We apply the unit conversion in the following manner: 
\begin{equation}
I = B(\lambda) \frac{10^9}{4\pi} \frac{hc}{\lambda} \simeq 0.00993 \times \frac{B(\lambda)}{2 \pi \lambda} \\
\end{equation}
where I is the spectral radiance in [W cm$^{-2}$ sr$^{-1}$ $\mu$m$^{-1}$], B is the surface brightness in [kR \AA$^{-1}$], $\lambda$ is the wavelength in [{\AA}], $h$ is Planck's constant, and $c$ is the speed of light.


Figure \ref{example} shows two spectra acquired by UVIS whilst Cassini was at local noon. The black line shows the equatorial reflectance spectrum, increasing in radiance towards longer wavelengths, whilst the orange line shows H$_2$ Lyman and Werner band emission excited by auroral electrons at the north pole, with a very small solar reflectance component. Both spectra contain strong emission from H Lyman-$\alpha$ emission, omnipresent about the Jupiter system \cite[e.g. see][]{doi:10.1029/GL015i010p01145, 2016Icar..278..238M}. There is evidence of very low level emission from H$_2$ in the equatorial spectrum, which is either due to some of the pixels containing polar emissions, or that there are a small population of electrons capable of exciting atomic hydrogen at low latitudes. The grey line in Figure \ref{example} shows the presence of so called {\it evil pixels}, which are pixels that have a response to illumination which is unpredictable, and are removed in this analysis. For a detailed description on the extraction and calibration of the Cassini UVIS PDS data see {\it Cassini UVIS User's Manual} contained within the \verb|couvis_0060| archive.


The uncertainty in the brightness observed by Cassini UVIS is calculated by the calibration routines provided on the PDS, and examples are provided in the User's Manual. The uncertainty on the radiance varies across the detector array, but does not vary as a function of the integration time, which one may expect if the errors are dominated by counting statistics. Here, we use observations with integration times of 60 and 67 seconds (see the following Sections). We assume that the errors from the calibration routine are appropriate for these exposure times, then propagate the errors when combining observations as appropriate. The resulting errors on the observed radiance appear sensible given the noise seen in the UVIS spectra. 

Here we will consider two set of UVIS observations with Cassini positioned at a) local noon at a distance of 250 R$_J$ and b) at closest approach to the planet at a distance of 137 R$_J$. The former provides a low-phase view of Jupiter, akin to what is observed from Earth, and the latter provides the highest spatial resolution view, albeit at high phase angles. The locations at which Cassini was at local noon and the time of closest approach to Jupiter are indicated as dots in Figure \ref{geometry}. Observations from these vantage points are detailed in the next two sections. 

\subsection{Local noon UVIS data \label{sec_obs_noon}}

On day 2000-348 Cassini was positioned at Jupiter's local noon, as illustrated in Figure \ref{geometry}. On the following days 2000-349 and 2000-350 UVIS acquired spectra at a distance of 250 R$_J$, with the slit aligned along the rotational axis. During this interval Jupiter subtended 5 spatial pixels. We added 167 individual spectral pixels obtained at full 1024 pixel spectral resolution, contained between -20 and +20 degrees latitude, from local-times between 11 and 13 for a total integration time of 2.7 hours (60 seconds per pixel), to produce a single FUV spectrum with a good S/N ratio of $\sim$20.



\begin{figure}
\centering
\includegraphics[width=1.0\textwidth]{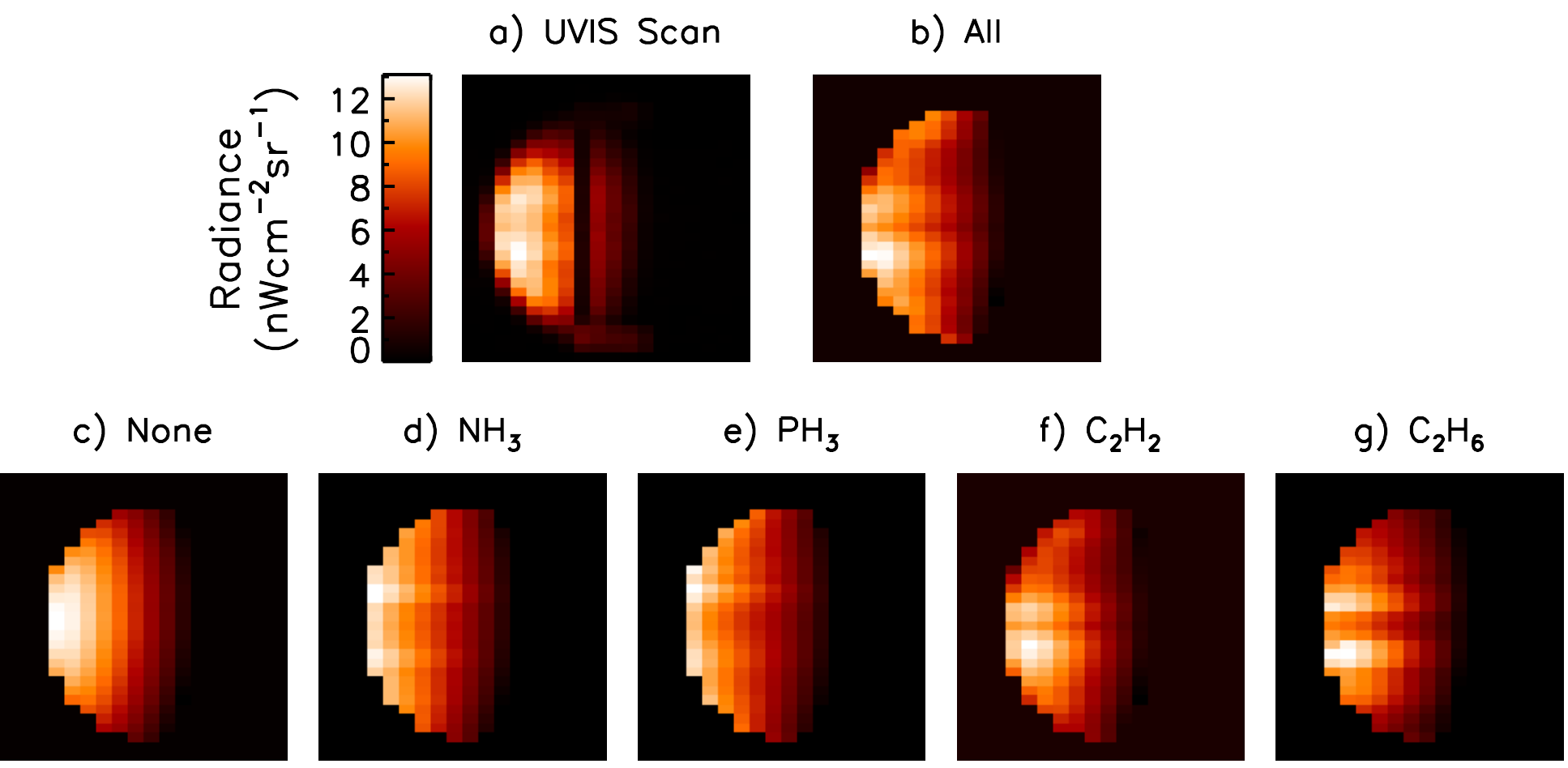}
\caption{ A comparison between the Cassini UVIS data obtained on 2001-001 02:53 and the NEMESIS forward models using the Cassini CIRS vertical profiles as input, with north being up. The top row shows the data and the forward model where scattering and absorption from all species is turned on, c) shows Rayleigh scattering only, with out any molecular absorption and d) to g) shows the scattering with only one species turned on, as indicated.  \label{scatcomp}}
\end{figure}

\subsection{High phase UVIS data}

At closest approach, at a distance of 137 R$_J$, the planet is contained within 24 spatial pixels. UVIS had the lowest spatial resolution of any of the remote sensing instruments onboard Cassini, but the closest approach time-frame provides an opportunity to examine if latitudinal differences in abundances can be inferred from the UVIS data. 

The slit was aligned along the equator of Jupiter, and the spacecraft slewed along the rotational axis to build up a complete scan of the disk of Jupiter. A number of scans were performed at 512 pixel reduced spectral resolution, as the spacecraft receded from the Jupiter system, but the first one obtained on 2001-001 provides the lowest phase view of the system, with an exposure time of 67 seconds per scan position. 

The scan of Jupiter is shown in Figure \ref{scatcomp}a, integrated in wavelength between 0.15 and 0.19 $\mu$m, so that each pixel represents an individual spectrum. The Sun is located towards the left in the Figure. The dark line in the middle of the scan is a pixel position that is poorly calibrated, and is excluded from this analysis. Both the northern and southern equatorial belts are visible and there is a clear solar angle dependence on the observed emission. The excited H$_2$ component of the auroral emission (at 0.16 $\mu$m, see Figure \ref{example}) is visible at southern pole in particular. 


\section{Results}

We have updated the NEMESIS radiative transfer code to model and retrieve atmospheric abundances from ultraviolet reflectance spectra, and we have identified a set of FUV observations with which we can test these model developments against. In this section we will apply the code to the Cassini UVIS Jupiter observations. First, we will examine how well NEMESIS forward models, using the available CIRS profiles, can reproduce the disc-resolved UVIS observations obtained during Cassini closest approach. Secondly, by retrieving atmospheric parameters from a set of synthetic spectra generated by adding random noise to forward models, we will determine how sensitive we are to abundances of different species. Thirdly, we calculate the two-way transmission function, which reveals what pressures in the atmosphere to which we are broadly sensitive. Lastly, we perform hypothesis testing to examine whether there are a set of atmospheric profiles that are consistent with both the CIRS and UVIS observations.

\subsection{Disk resolved forward models}

As a first step, we ask: how well can we reproduce the UVIS observations using the vertical profiles of temperature and abundance that CIRS derived as a function of latitude? 

The spatially-resolved UVIS observations obtained during closest approach can be seen in Figure \ref{scatcomp}a. We generate a set of forward models with different molecular absorptions turned on, based on the UVIS spectral range and spectral resolution, specifying the phase, emission and incidence angle of each pixel. The CIRS profiles form the input atmosphere, using the set of vertical profiles appropriate for each latitude. The wavelength integrated output of these forward models are seen in Figure \ref{scatcomp}. 

Figure \ref{scatcomp}c shows Rayleigh scattering only, rendering Jupiter as a smooth sphere. This uniformity shows that the scattering is independent of temperature, since the CIRS profiles have temperatures that vary with latitude \cite[e.g.][]{2016Icar..278..128F}. Figure \ref{scatcomp}d shows Rayleigh scattering combined with ammonia absorptions, showing bright bands below and above the equator. Here, a bright band indicates less absorption and less ammonia, and so ammonia abundances are higher at the equator than at mid-latitudes. The phosphine absorption is shown in Figure \ref{scatcomp}e, with some minor latitudinal depletions, particularly at northern mid-latitudes. Figure \ref{scatcomp}f and \ref{scatcomp}g shows the absorption by acetylene and ethane respectively, both showing asymmetric abundances, with the latter being integrated between 0.154 and 0.156 $\mu$m. The bright SEB indicates lower abundances for both species. Finally, Figure \ref{scatcomp}b shows the combination of Rayleigh scattering and absorption from all four species, replicating the UVIS observations in Figure \ref{scatcomp}a relatively well. In particular, the forward models manages to reproduce the observed asymmetry between the NEB and SEB radiances. There are more significant differences at higher latitudes, however. The scattering by stratospheric aerosols becomes greater at these locations. Since this is not encapsulated in the forward models we make no attempt to include the aerosols, resulting in an overestimation of the radiance. 

The data and forward models in Figure \ref{scatcomp} qualitatively show that there is significant information about atmospheric abundances of acetylene, ethane, ammonia and phosphine contained within the UVIS wavelength range. In the coming section we will quantify the extent to which these abundances can be retrieved from UVIS observations.

\begin{figure}
\centering
\includegraphics[width=0.9\textwidth]{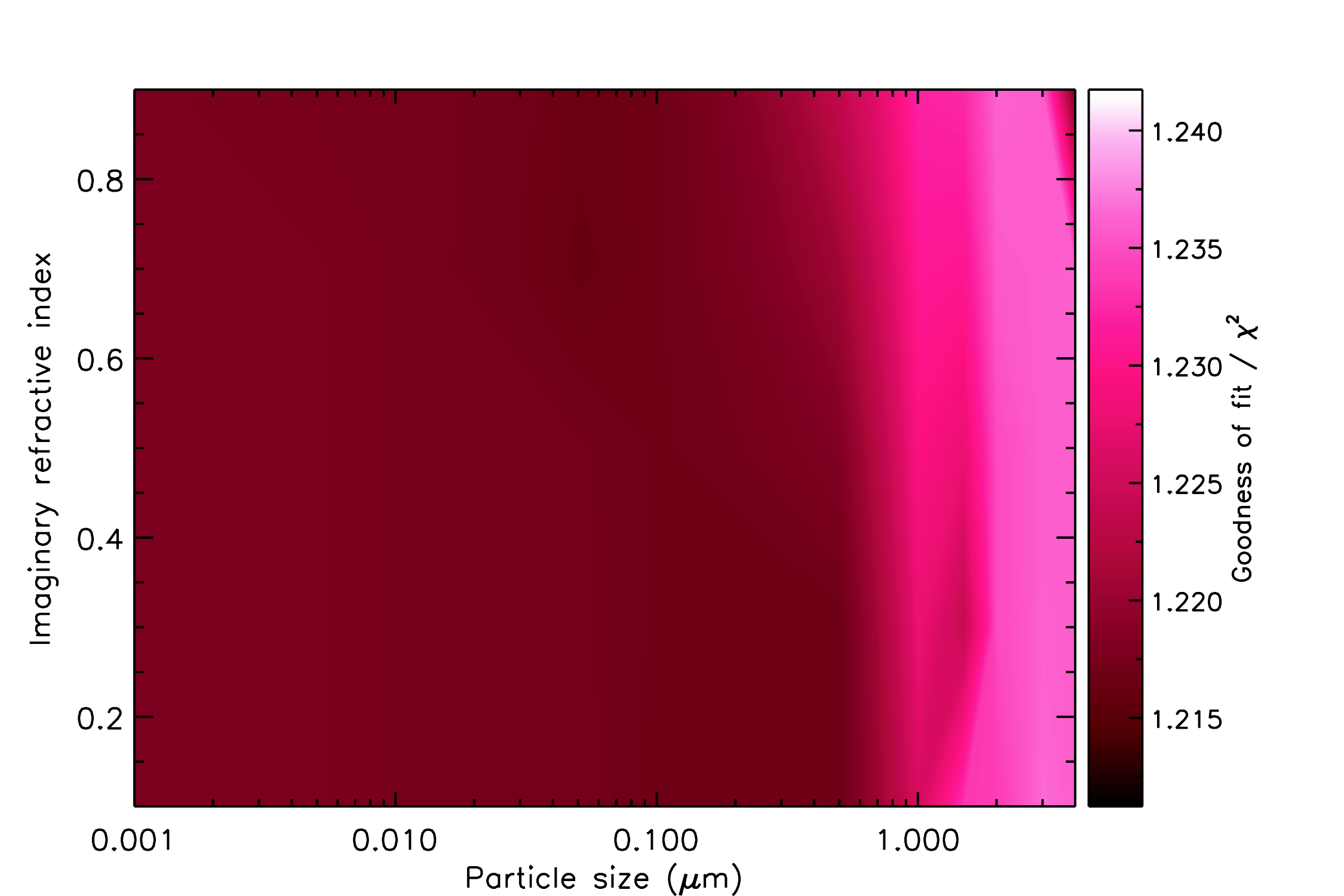}
\caption{The retrieved goodness of fit, $\chi^2$, as a function of aerosol imaginary refractive index and particle size. In these retrievals, the scaled aerosol abundances are retrieved as to achieve a good fit to the observed spectrum. The $\chi^2$ only varies less than 2\%, indicating that the ultraviolet is insensitive to these two parameters.  \label{ars}}
\end{figure}

\subsection{Sensitivity to aerosols \label{aersec}}

In the ultraviolet, aerosols act as to absorb and scatter the light away from and towards the observer, potentially significantly reducing the observed radiance. To investigate whether we are sensitive to any particular aerosol property we devise a sensitivity test, retrieving scaled vertical abundances of acetylene, ethane, phosphine, ammonia, and aerosols for a range of different aerosol properties. Figure \ref{ars} shows the retrieved goodness-of-fit parameter, $\chi^2$, for a range of imaginary refractive indices and aerosol particle sizes between 0.001 and 3 $\mu$m, with a base pressure of 10 mbar and a fractional scale height of 1. Despite this large parameter space, there is very little change in the $\chi^2$, with a maximum change of 2\%. Consequently, whilst we require some form of aerosol distribution to achieve good retrievals, particularly at high latitudes (e.g. Figure \ref{scatcomp}a), we cannot determine the exact physical properties of the aerosols. However, as we will show in the following Sections, the simple assumptions about the aerosol properties outlined in Section 2.3 are sufficient to produced good fits to the observed Cassini UVIS spectrum: a single particle size of 0.3 $\mu$m, refractive index of $1.4 + 0.3i$, and an optical depth of one at 1 bar, with an aerosol base pressure of 10 mbar and a fractional scale height of 1. 



\begin{figure}
\centering
\includegraphics[width=0.9\textwidth]{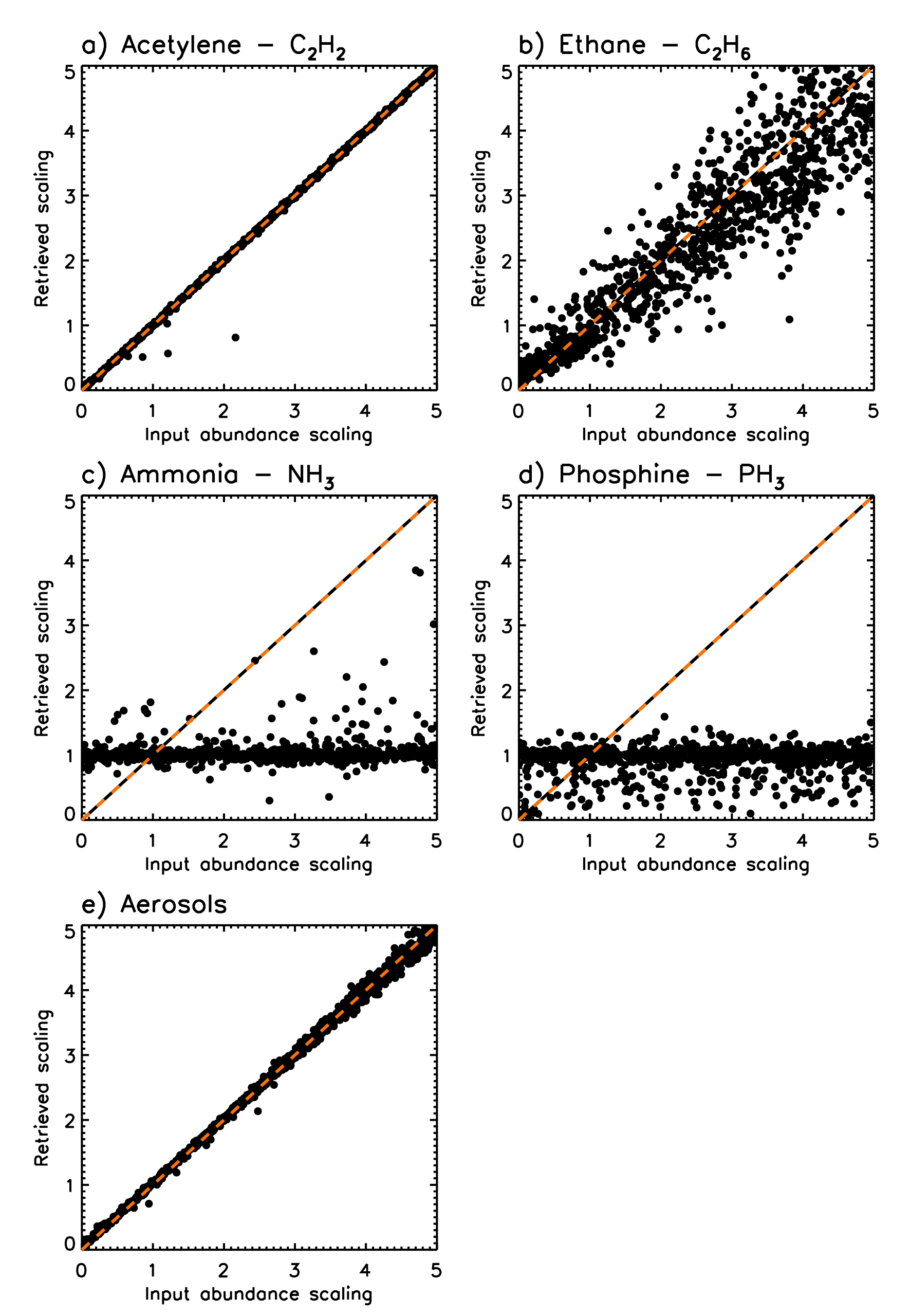}
\caption{A retrieval sensitivity study for the Cassini UVIS FUV channel with wavelength coverage up to 0.19 $\mu$m, showing the input abundance scaling of the vertical profiles versus the scaling retrieved using NEMESIS, as described in Section \ref{uvisretsec}. We are unable to retrieve ammonia and phosphine, but can with confidence retrieve acetylene and the aerosol abundance, and to a lesser extent the ethane abundance. The dashed line shows the expected trend when the sensitivity to retrieve a particular species is optimal. \label{sensuvis}}
\end{figure}

\subsection{Cassini UVIS retrievability study \label{uvisretsec}}

Figure \ref{scatcomp} shows that there are absorption features of acetylene, ethane, ammonia, and phosphine present in the forward models. However, the goal of this study is to retrieve these abundances from observed UVIS spectra, and the degree to which we are able to do so is determined by an interplay between the Rayleigh-scattered solar spectrum, the different molecular absorption features, aerosol scattering, and, most importantly, the S/N of the observed spectrum. To test the degree to which we can retrieve abundances from UVIS spectra, we devise a sensitivity study. 



We generate 1,000 NEMESIS forward models in the UVIS spectral range and at the same spectral resolution, using the equatorial CIRS profiles (see Figure \ref{apriori}) as a starting atmosphere. For each, the vertical abundance profiles of acetylene, ethane, ammonia, phosphine, and aerosols are each scaled by random values between 0 and 5 for each forward model. The output of these forward models represent idealistic observations, and to simulate real data, we add noise across each spectrum equivalent to the S/N of a typical UVIS observation. We then retrieve the abundance scaling using NEMESIS. The correlation between the original scaling and the retrieved scaling then gives us a measure of the retrievability of each species. This is shown in Figure \ref{sensuvis}. 

When we have good sensitivity to the abundance of a particular species, the retrieved scaling will be the same as the input scaling, and the 1,000 retrievals will form a diagonal line, shown as a dashed line in Figure \ref{sensuvis}. Here, we see we are able to achieve excellent sensitivity to acetylene (Figure \ref{sensuvis}a), with the retrieved values falling on to the diagonal line, with an error on the retrieved scaling of $\sim2\%$. We are only moderately sensitive to ethane (Figure \ref{sensuvis}b), driven primarily by the fact that the ethane feature is confined to a narrow wavelength region (A in Figure \ref{fmadd}a), producing an error of $\sim20\%$ on the retrieved scalings. Figure \ref{sensuvis}c and \ref{sensuvis}d shows the sensitivity for ammonia and phosphine, both having a flat distribution of retrieved scalings hovering about one, both with an error of $\ge100\%$. This means that UVIS is effectively blind to variations these species, unable to provide the required contrast. The aerosol sensitivity in Figure \ref{sensuvis}e shows an excellent agreement between the input and retrieved scaling with a scaling error of $\sim2.5\%$. However, since there is degeneracy between many different aerosol properties (e.g. particle size, optical depth, altitude), the aerosol scaling does not offer any physical insight to their specific properties, only that they attenuate the radiance uniformly across the UVIS spectral range. 


\begin{figure}
\centering
\includegraphics[width=0.9\textwidth]{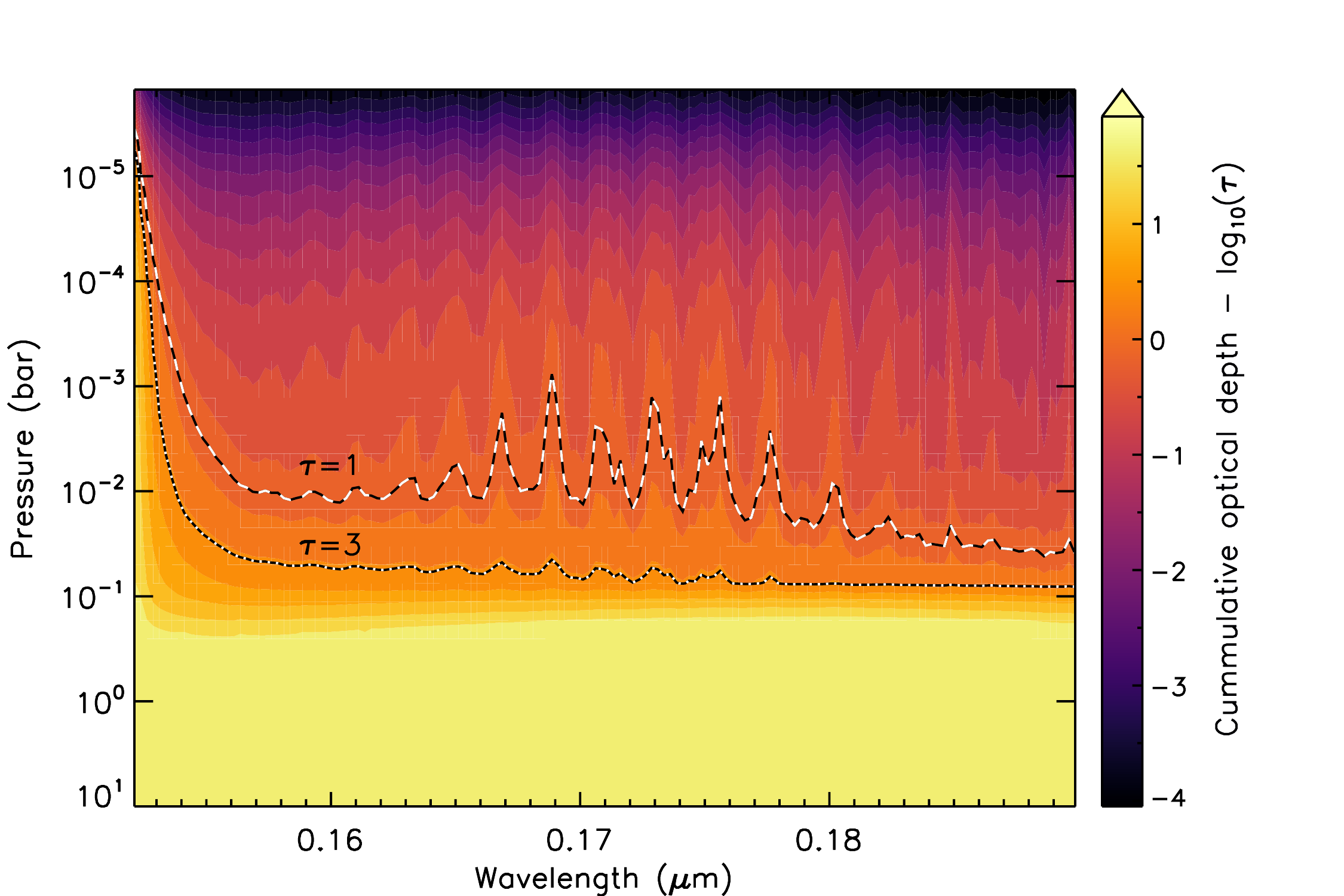}
\caption{The cumulative extinction optical depth of the Cassini UVIS spectrum as a function of wavelength and pressure, as calculated from the two-way transmission function. This includes Rayleigh scattering and absorption from acetylene, ethane, ammonia, and phosphine. The dashed lines shows where the optical depth equals unity, and the dotted line shows where the optical depth is 3. \label{oduvis}}
\end{figure}

\begin{figure}
\centering
\includegraphics[width=0.9\textwidth]{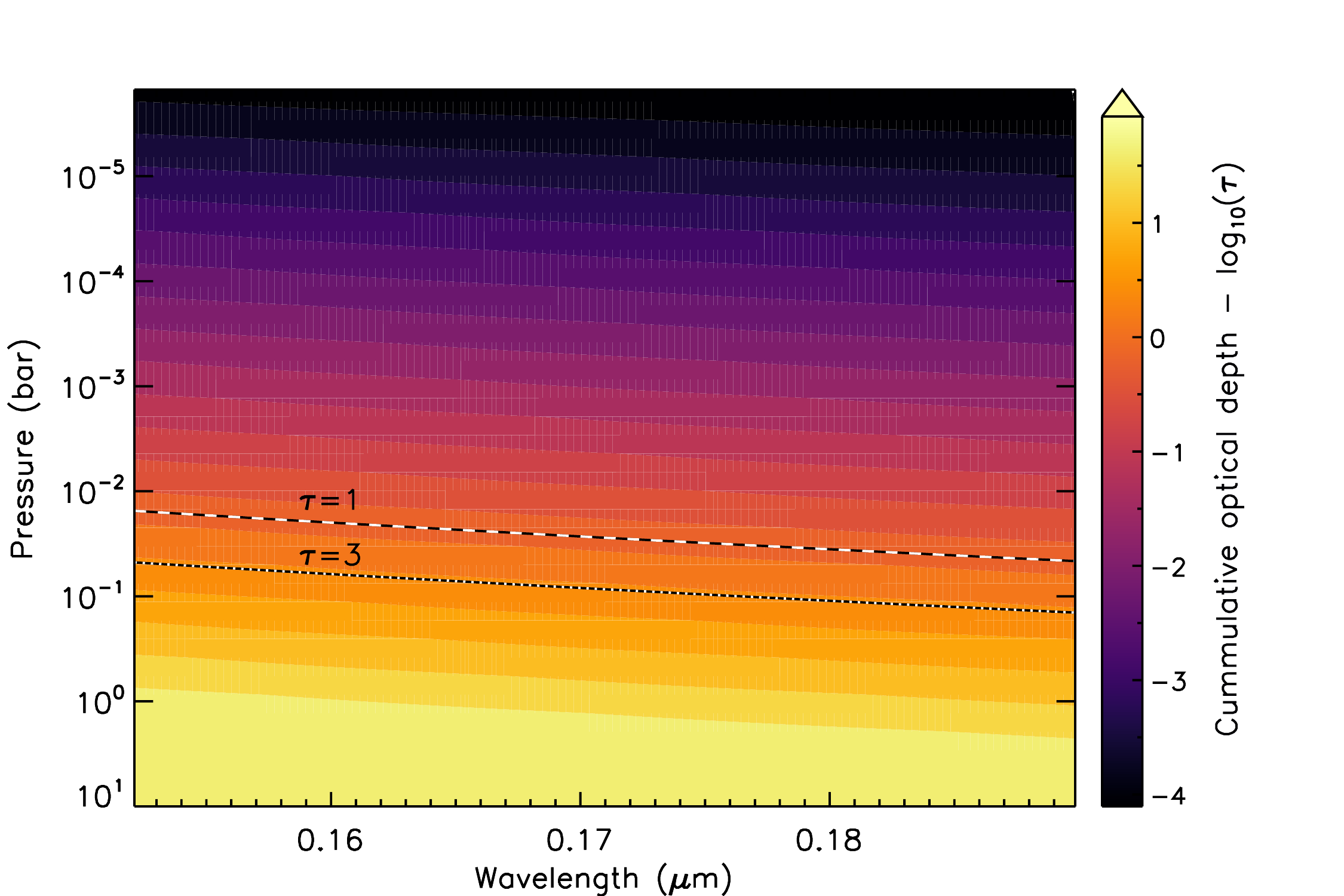}
\caption{The optical depth of Rayleigh scattering only in the Cassini UVIS wavelength range, with the ultraviolet molecular absorptions turned off. The dashed lines traces the pressure levels where the optical depth is unity. Comparing this to Figure \ref{oduvis} shows that the molecular absorptions act as to increase the altitude of the unity optical depth, and therefore the altitude of the peak instrument sensitivity. \label{rayonly}}
\end{figure}

\subsection{Optical depth of the UVIS spectrum \label{odsect}}

The previous section demonstrated that we can retrieve acetylene, ethane abundances, and aerosol attenuation from Cassini UVIS spectra. What pressure levels in the atmosphere of Jupiter is Cassini UVIS we sensitive to? The optical depth, shown in Figure \ref{oduvis}, can be calculated as the natural logarithm of the two-way transmission function for aerosol free conditions, which in turn can be generated by the NEMESIS code. Note that this calculation does not include multiple scattering, in which case $\tau = 1$ provides the altitude of the peak sensitivity. However, if incoming photons undergo multiple scattering, governed by the single-scattering albedo, then they can penetrate to lower altitudes before being absorbed. In addition, some of them will scatter back out of the atmosphere along the line-of-sight of the observer, providing a source to offset absorptions in the atmosphere. Therefore, in the presence of multiple scattering the $\tau = 1$ level in Figure \ref{oduvis} provides a measure of the top of the region that the ultraviolet photons originate from, with the lower bound being at lower altitudes, approximately $\tau \sim 3$. A more robust way to determine the range of altitudes that UVIS is sensitive is to determine the contribution function, which is not trivial to extract and is dependent on the model chosen for the aerosol absorption and scattering, and is outside the scope of the present study. 



The optical depth reaches unity at around 10 mbar, and there are peaks throughout the spectral range at which the atmosphere becomes increasingly opaque at pressures less than 1 mbar. These peaks are produced by corresponding peaks in the acetylene absorption cross-section, producing spectral regions with a peak sensitivity at around 1 mbar.  Similarly, the ethane absorption slope is located around 0.155 $\mu$m (see Figure \ref{fmadd}a), where the optical depth reaches unity at $\sim$5 mbar, representing the top of the altitude sensitivity of UVIS. 

Figure \ref{oduvis} shows that wavelength regions with large molecular absorption cross-sections sample higher altitudes in the atmosphere. However, what altitudes does pure Rayleigh scattering sample? To test this, we generated the two-way transmission function as before, but with the ultraviolet molecular absorptions turned off - this is shown in Figure \ref{rayonly}. Rayleigh scattering samples pressures less than tens of mbar, and moves to greater pressures with increasing wavelengths. At 0.19 $\mu$m the optical depth reaches unity at $\sim$80 mbar, comparing favourably to previous estimations \citep[e.g.][]{1974ApJ...187..641T}.

The optical depth calculations performed here do not include the effect of scattering by aerosols. In Section \ref{aersec} we showed that the ultraviolet wavelength range is largely insensitive to the physical properties of the aerosol population, and that they act to uniformly reduce the radiance of the observed spectrum, unless placed high enough (above the absorbing layer) in large abundances. Changing the aerosol distribution will potentially have a significant effect on the optical depth of these observations. 



\begin{figure}
\centering
\includegraphics[width=0.9\textwidth]{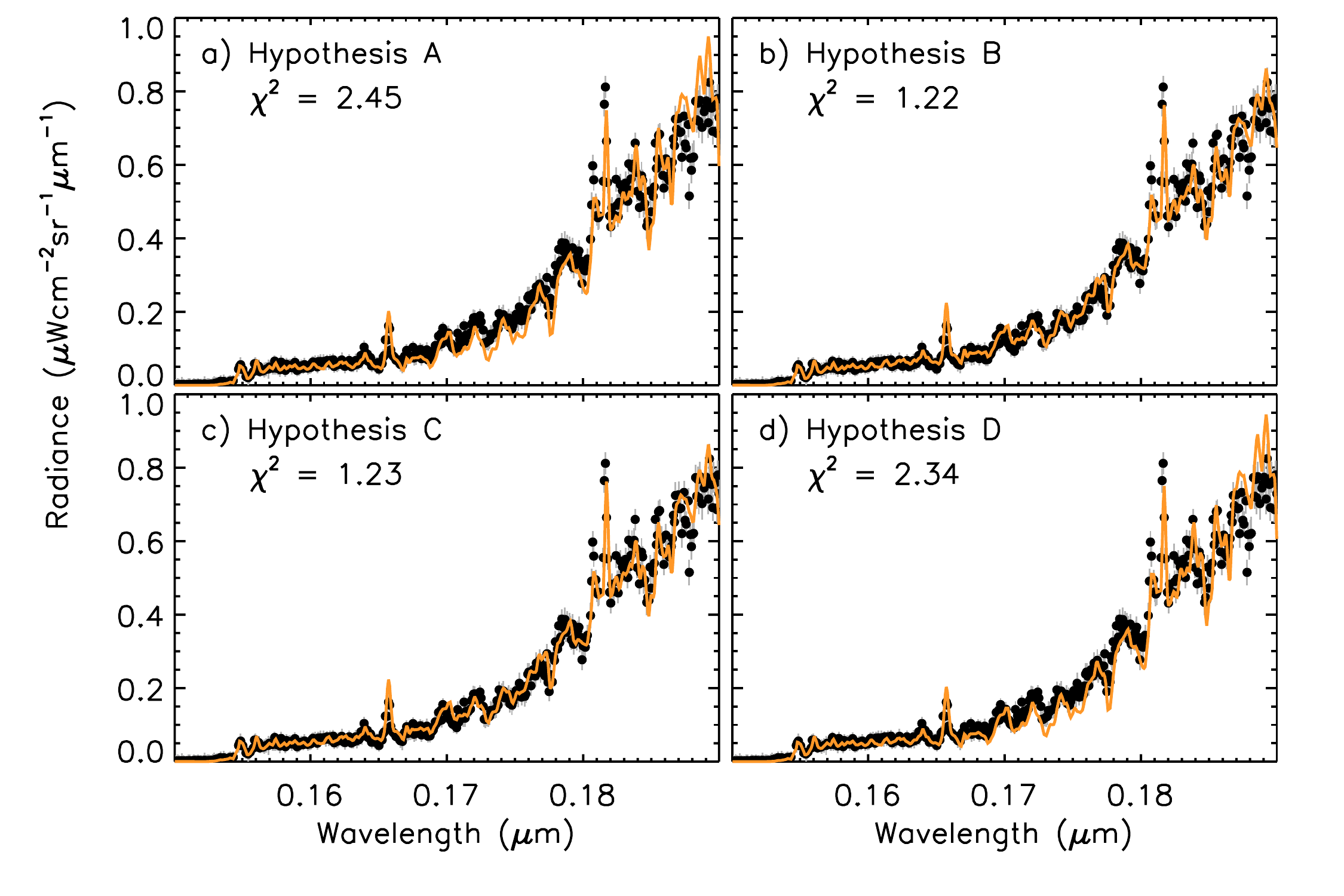}
\caption{The spectrum observed by Cassini UVIS at local noon, centred at the equator, with the results of the four Hypotheses: A, B, C, and D, and the resulting $\chi^2$ goodness-of-fit. \label{noon_spectrum}}
\end{figure}


\subsection{Hypothesis testing}

Since UVIS and CIRS are both able to retrieve the abundance of acetylene and ethane in the stratosphere of Jupiter, are the retrieved abundances from the two instruments consistent with each other? To test this, we postulate four testable hypotheses: 

\begin{itemize}
\item {\bf Hypothesis A: } The abundance profiles of acetylene and ethane derived from CIRS observations are sufficient to fully reproduce the UVIS observations. Since the aerosol population captured in the ultraviolet are not captured by the CIRS data, we here only retrieved scaled aerosol abundances. 

\item {\bf Hypothesis B: } The observed UVIS spectra are incompatible with the CIRS abundance profiles. In order to determine what difference in abundance is required to better fit the UVIS spectra, we retrieve scaled acetylene, ethane, and aerosol abundances. 

\item {\bf Hypothesis C: } There exists is a set of vertical profiles that are consistent with both UVIS and CIRS observations, with the two instruments sampling acetylene and ethane at different pressures levels in the atmosphere. In this manner, combining the two instruments provides a more complete view of stratospheric acetylene and ethane during the Cassini encounter. Here we retrieve complete vertical profiles of acetylene and ethane, but fixing the abundances at pressures $>$1 mbar -- levels that the CIRS observations are sensitive to.

\item {\bf Hypothesis D: } Same as C, but allowing acetylene and ethane only to change at pressures larger than 30 mbar. This provides an opportunity to examine the sensitivity of UVIS at altitudes lower than those to which CIRS is sensitive. 

\end{itemize}

In the following sections, these four tests will be applied to both the UVIS noon and closest-approach data.

\subsection{Observations at local noon \label{sectype}}


Figure \ref{noon_spectrum} shows the UVIS spectrum of Jupiter obtained at local noon, centred on the equator. The Figure \ref{noon_spectrum}a shows the result from retrieving only scaled aerosol abundances, with the acetylene and ethane abundance fixed at those derived from CIRS observations -- Hypothesis A, producing a $\chi^2 = 2.45$. Qualitatively, this fit is not excellent, producing an acetylene absorption that is much too deep between 0.17 and 0.18 $\mu$m. This shows that the retrieved CIRS abundances are too high for acetylene to explain the UVIS observations. In contrast, the model reproduces the ethane feature at 0.155 $\mu$m well.

If the data-sets from the two wavelengths regimes are inconsistent, then what fraction of the CIRS abundances of acetylene and ethane are required to reproduce the UVIS observations? To answer this question -- Hypothesis B -- we retrieve the aerosol scaling, as before, in addition to the multiplicative abundance scaling factor for acetylene and ethane. This is shown in Figure \ref{noon_spectrum}b, producing a $\chi^2 = 1.22$. Hypothesis B produces a much better fit than Hypothesis A, capturing the shallower absorption features of both acetylene and ethane. The abundance scaling factors are $0.49 \pm 0.03$ for acetylene and $1.1 \pm 0.3$ for ethane, producing abundances at 5 mbar of $48 \pm 3$ ppb and $3.7 \pm 0.9$ ppm respectively. 

The scaled retrieval of Hypothesis B produces a good fit to the data, but since the near simultaneous CIRS observations in the mid-infrared already constrain the abundance of acetylene and ethane at pressures $>$1 mbar, the scaled retrievals are unlikely to be physical. Figure \ref{oduvis} suggests that the UVIS spectra may be sensitive to lower pressure levels than the CIRS observations. To generate vertical profiles of acetylene and ethane that are consistent with both UVIS and CIRS, we test Hypothesis C, fixing the retrievals at pressures $>$1 mbar to those retrieved from CIRS observations. This is shown in Figure \ref{noon_spectrum}c, and has a $\chi^2 = 1.23$, providing a near-identical goodness-of-fit as Hypothesis B. We retrieve abundances at 0.1 mbar of $1.21 \pm 0.07$ ppm and $20.8 \pm 5.1$ ppm for acetylene and ethane respectively. 

Hypothesis C provides a set of acetylene and ethane profiles that are compatible with both CIRS and UVIS observations by altering abundances at pressures lower than 1 mbar. Is it possible to achieve similarly consistent profiles by varying the abundances {\it below} the altitude at which CIRS is sensitive to? This is examined by Hypothesis D, where scaled aerosols are retrieved as well as a full retrieval of acetylene and ethane at pressures larger than 30 mbar, the low altitude end of the CIRS sensitivity \citep{2007Icar..188...47N}. This  is shown in Figure \ref{noon_spectrum}d, and the resulting fit to the data is poor, showing that we are unable to provide a set of profiles that are consistent with both CIRS and UVIS by altering abundances at pressures $>$30 mbar, indicative that UVIS has no or limited sensitivity at these altitudes. 

\begin{figure}
\centering
\includegraphics[width=0.9\textwidth]{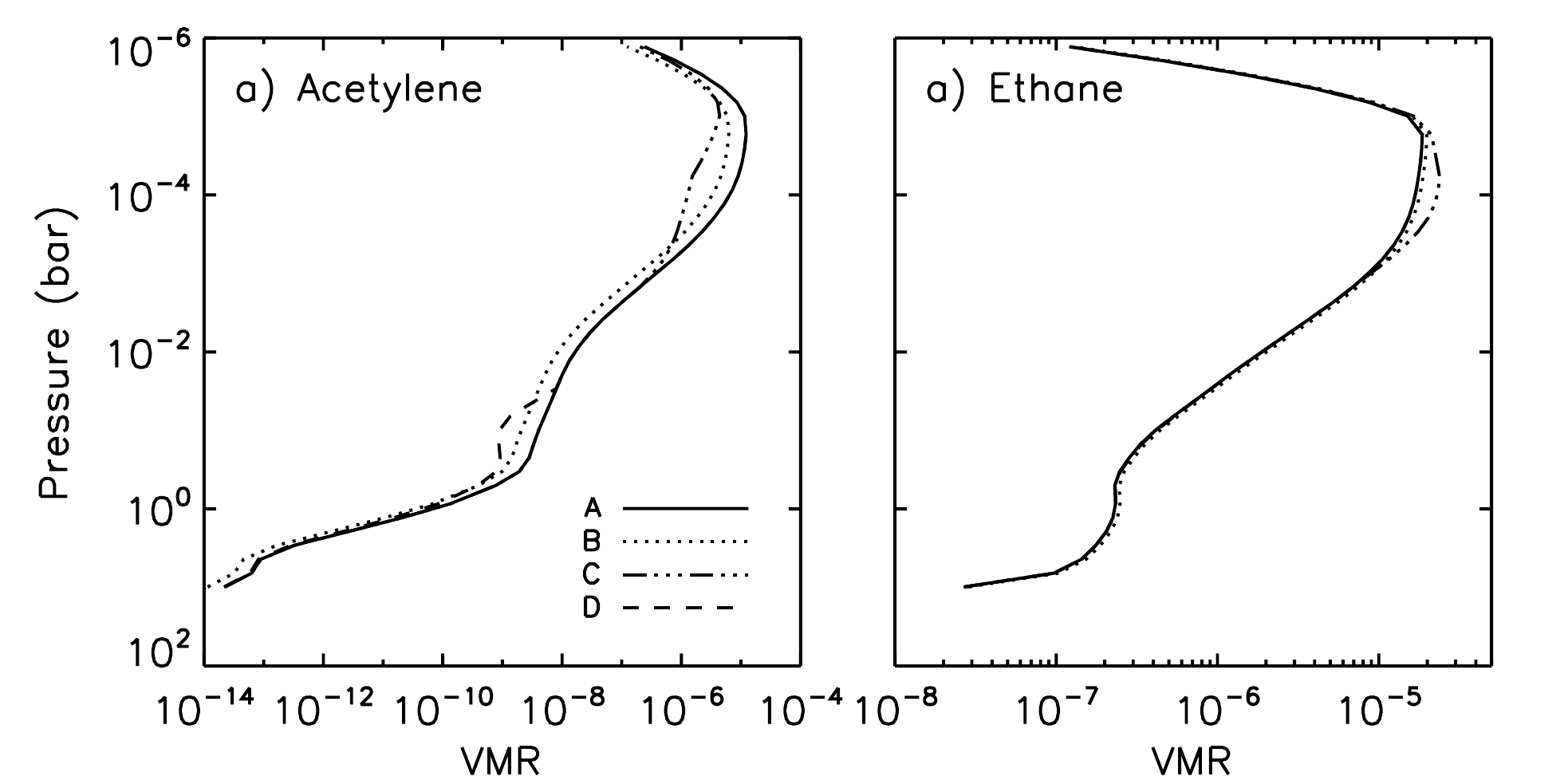}
\caption{The volume mixing ratio (VMR) of acetylene and ethane as a function of pressure for the equatorial UVIS noon observations, showing the profiles derived from Hypothesis A, B, C, and D for the low phase noon observations, as described in Section \ref{sectype}. 
 \label{noon_profiles}}
\end{figure}

The vertical profiles of acetylene and ethane for the four hypotheses are shown in Figure \ref{noon_profiles}. The dot-dot-dashed profiles of Hypothesis C changes the abundances at high altitudes, which produces a near identical goodness-of-fit to the UVIS noon spectrum as Hypothesis B, meaning that the profiles of Hypothesis C are an equally valid set of profiles, but are also consistent with the CIRS observations. The retrieval of Hypothesis D results in a deep excursion form the prior (solid) of acetylene at pressures $>$30 mbar, yet this is insufficient to produce a good fit to the data -- see Figure \ref{noon_spectrum}d. The vertical profile of Hypothesis D for ethane shows no deviation from the prior at all, indicating that UVIS observations have no sensitivity to ethane at all at pressures $>$30 mbar. 


\begin{figure}
\centering
\includegraphics[width=0.9\textwidth]{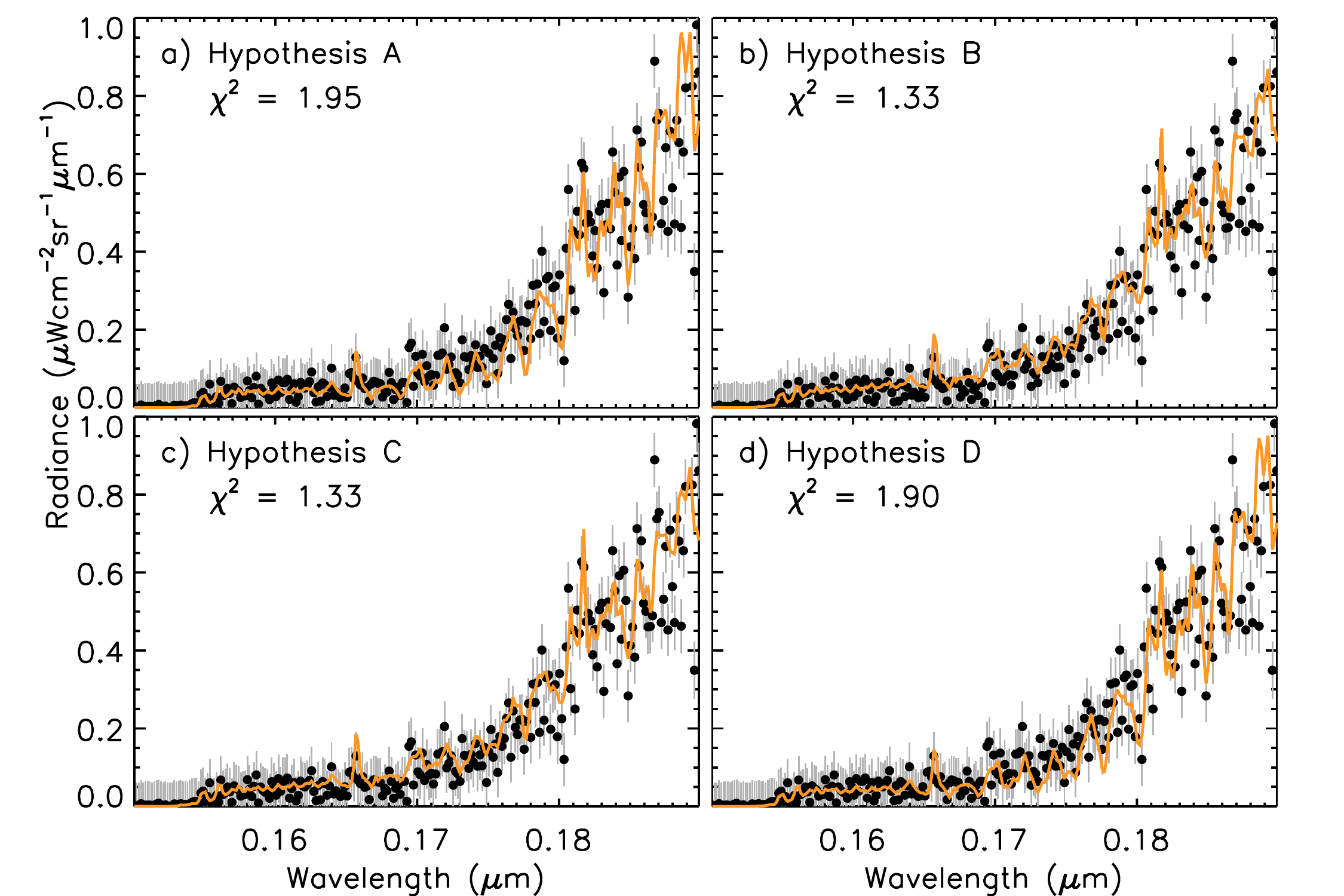}
\caption{The Cassini UVIS observation of the NEB obtained during Cassini's closes approach to Jupiter (dots), with NEMESIS spectrum for retrievals A, B, C, and D as described in Section \ref{sectype}, shown as solid orange lines.  \label{ca_spectrum}}
\end{figure}

\subsection{Spatially resolved observations at closest approach}

The UVIS observations obtained at closest approach provide the opportunity to retrieve spatially resolved latitude profiles of acetylene and ethane from ultraviolet reflectance spectra. The observed spectrum obtained at the NEB can be seen as dots in Figure \ref{ca_spectrum}, and has a significantly lower S/N than the combined spectra obtained at local noon, because each spatial position is here an individual observation of 67 seconds. Here, we perform the same four hypothesis tests: A, B, C, and D. 

The result of testing Hypothesis A in Figure \ref{ca_spectrum}a. It has a $\chi^2 = 1.95$, whilst all latitudes have a median $\chi^2 = 2.09$. Similarly to Hypothesis A in Figure \ref{noon_spectrum}a, this spectrum has deep absorption features that are inconsistent with the data, particularly about 0.17 to 0.18 $\mu$m range, where acetylene is a strong absorber. 

\begin{figure}
\centering
\includegraphics[width=0.9\textwidth]{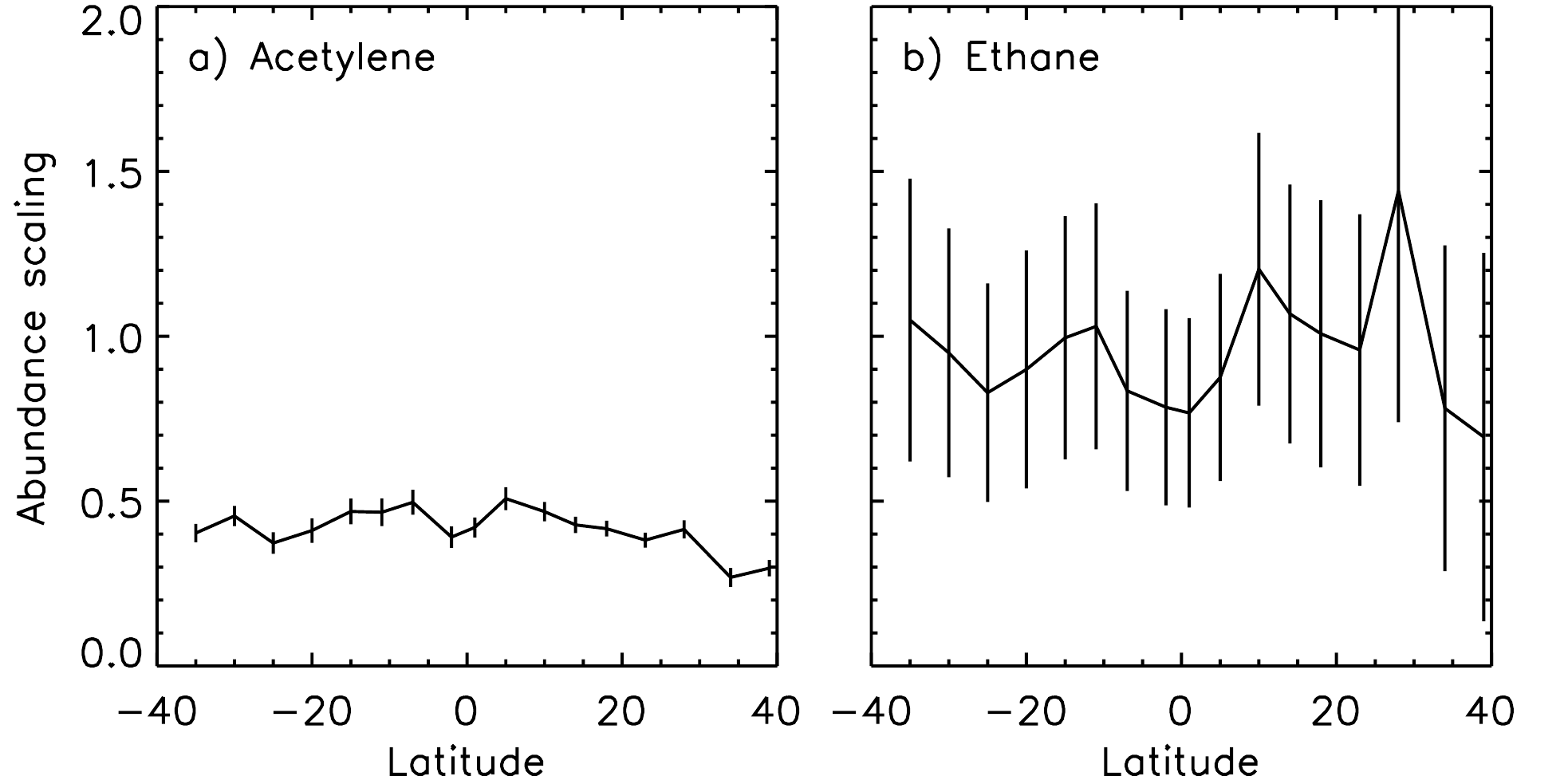}
\caption{The acetylene and ethane abundance scaling factors retrieved for Hypothesis B from the closest approach Cassini UVIS data as a function of latitude. \label{ca_scalings}}
\end{figure}

The NEB spectrum for Hypothesis B is shown in Figure \ref{ca_spectrum}b, with a retrieval $\chi^2 = 1.33$, with a median of $\chi^2 = 1.41$ over all latitudes. The retrieved acetylene and ethane scaling factors are shown in Figure \ref{ca_scalings}a as a function of latitude, with a mean of $0.42\pm 0.06$ and $0.95 \pm0.18$, respectively. This translates to 5 mbar abundances of $57 \pm 3$ ppb for acetylene and $3.4 \pm 1.4$ ppm at the NEB. The acetylene scalings have a distinct minimum at the equator, with peaks at about $\pm7^{\circ}$ latitude, falling off approximately symmetrically towards the pole. The error bars of the ethane scaling factors are significant, however, and there are three distinct minima, two at $\pm 20^{\circ}$ and about the equator. 

The retrieval of Hypothesis C at the NEB is shown in Figure \ref{ca_spectrum}c, and has a $\chi^2 = 1.33$, identical to Hypothesis B, showing a good fit to the observations. Hypothesis D, shown in Figure \ref{ca_spectrum}d provides a poorer fit to the data than both Hypothesis B and C, but one that is slightly better than Hypothesis A. 

\begin{figure}
\centering
\includegraphics[width=0.9\textwidth]{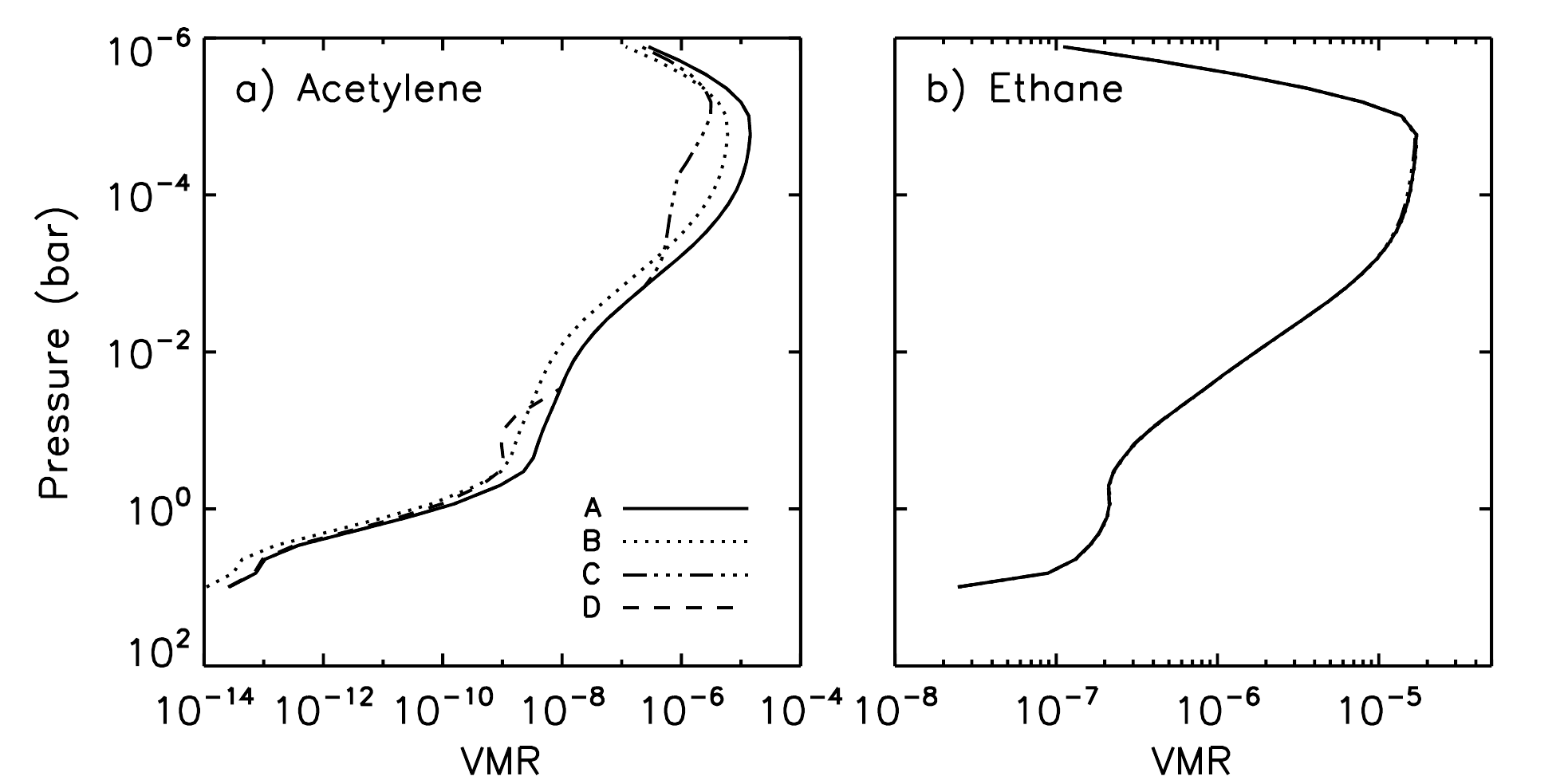}
\caption{The volume mixing ratio (VMR) of acetylene and ethane at the NEB as a function of pressure, showing the profiles derived from retrievals A, B, C, and D, as described in Section \ref{sectype}. 
 \label{ca_profiles}}
\end{figure}

The retrieved vertical profiles for all four hypotheses at the NEB can be seen in Figure \ref{ca_profiles}, which are very similar to those in Figure \ref{noon_profiles}. Hypothesis C produces reduced abundances at high altitudes for both species, whilst keeping the lower altitudes consistent with the CIRS observations. The 0.1 mbar abundances at the NEB are $0.78 \pm 0.05$ ppm for acetylene and $15.3 \pm 6.1$ ppm for ethane.

\subsection{CIRS compatibility}

\begin{figure}
\centering
\includegraphics[width=0.9\textwidth]{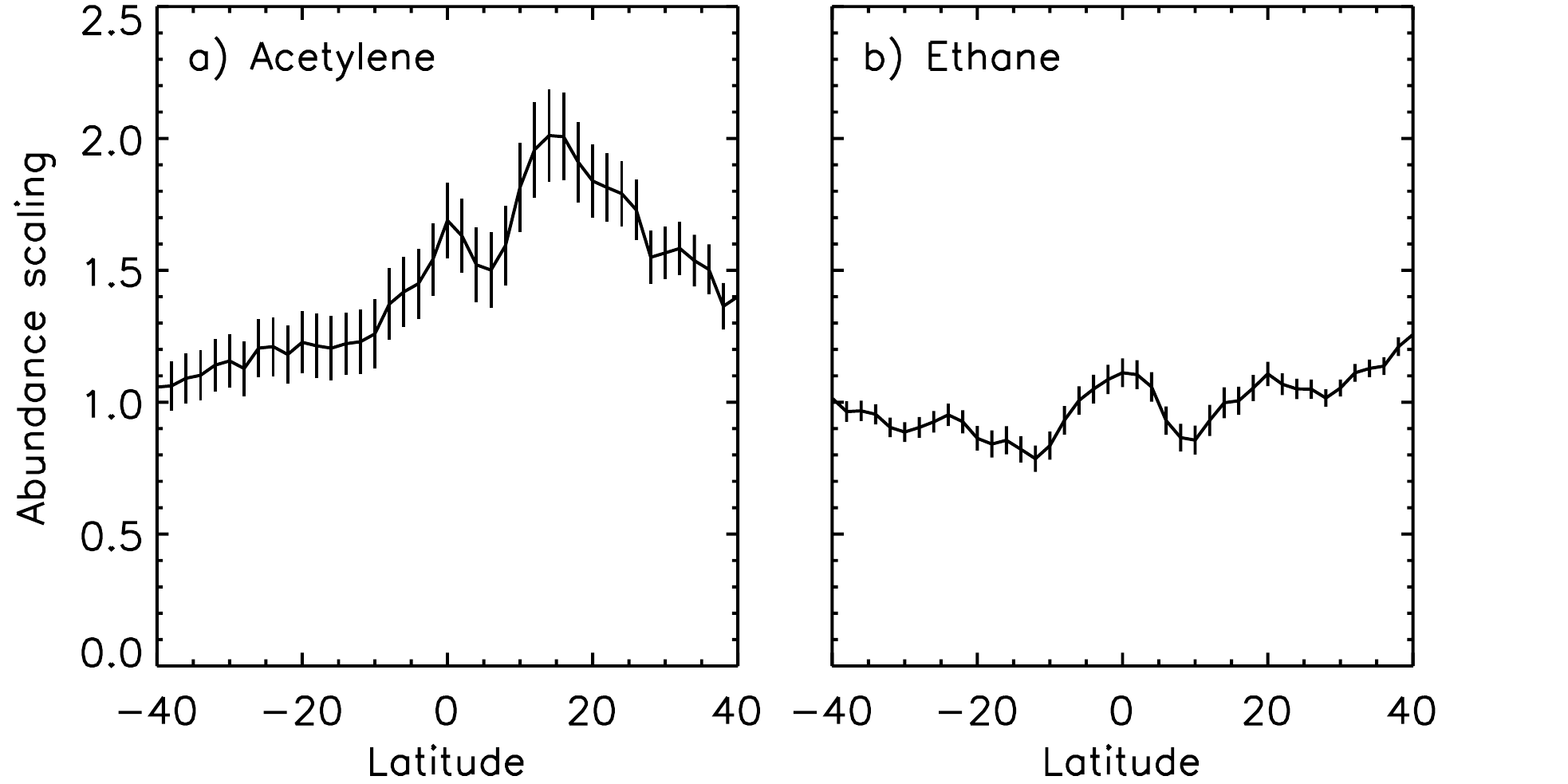}
\caption{The acetylene and ethane abundance scaling factors retrieved from Cassini CIRS data as a function of latitude, derived from using the new vertical profiles of Hypothesis C in Figure \ref{ca_scalings}. \label{cirs_scalings}}
\end{figure}

Hypothesis C retrieves vertical abundance profiles of acetylene and ethane that are the same as the CIRS prior at pressures $>$1 mbar. Are the new profiles in Figure \ref{ca_profiles} still consistent with the original CIRS observations? To test this we use the average acetylene and ethane profiles retrieved from the closest approach data for Hypothesis C as the new {\it a-priori} atmosphere and retrieve scaled acetylene and ethane abundances from the CIRS observations. The CIRS data and the procedure for retrieving abundances using NEMESIS from CIRS spectra are described in detail by \cite{2016Icar..278..128F}. The retrieved scaling factors as a function of latitude are seen in Figure \ref{cirs_scalings}, and there is only a very small degradation in the model-data $\chi^2$ compared to the original \cite{2016Icar..278..128F} retrievals, with a change of less than 1\%. The scaling factors reproduce the latitudinal structures retrieved by  \cite{2016Icar..278..128F} very well. This shows that it is possible to fit the CIRS observations with the new priors of Hypothesis C of Figure \ref{ca_profiles}, and that the CIRS observations are broadly compatible with reduced abundances at high altitudes. 

The abundance scalings in Figure \ref{cirs_scalings} show the same latitudinal features as described in \cite{2016Icar..278..128F}, with an acetylene enhancement about the NEB, and an ethane enhancement about the equator. At the equator and the NEB, the scaling factor reaches beyond 1.5, which is less than the original factor of $\frac{1}{0.4} = 2.5$ required to scale the original CIRS profiles to match the UVIS observations. Elsewhere, the scaling factor is slightly larger than one, meaning that the new {\it a-priori} profile is largely consistent with both the CIRS and the UVIS acetylene observations. 

In Figure \ref{cirs_scalings}b we show the ethane scalings which are very close to unity, indicating that they new a-priori is appropriate for use for use when analysing both UVIS and CIRS observations.



\begin{figure}
\centering
\includegraphics[width=0.9\textwidth]{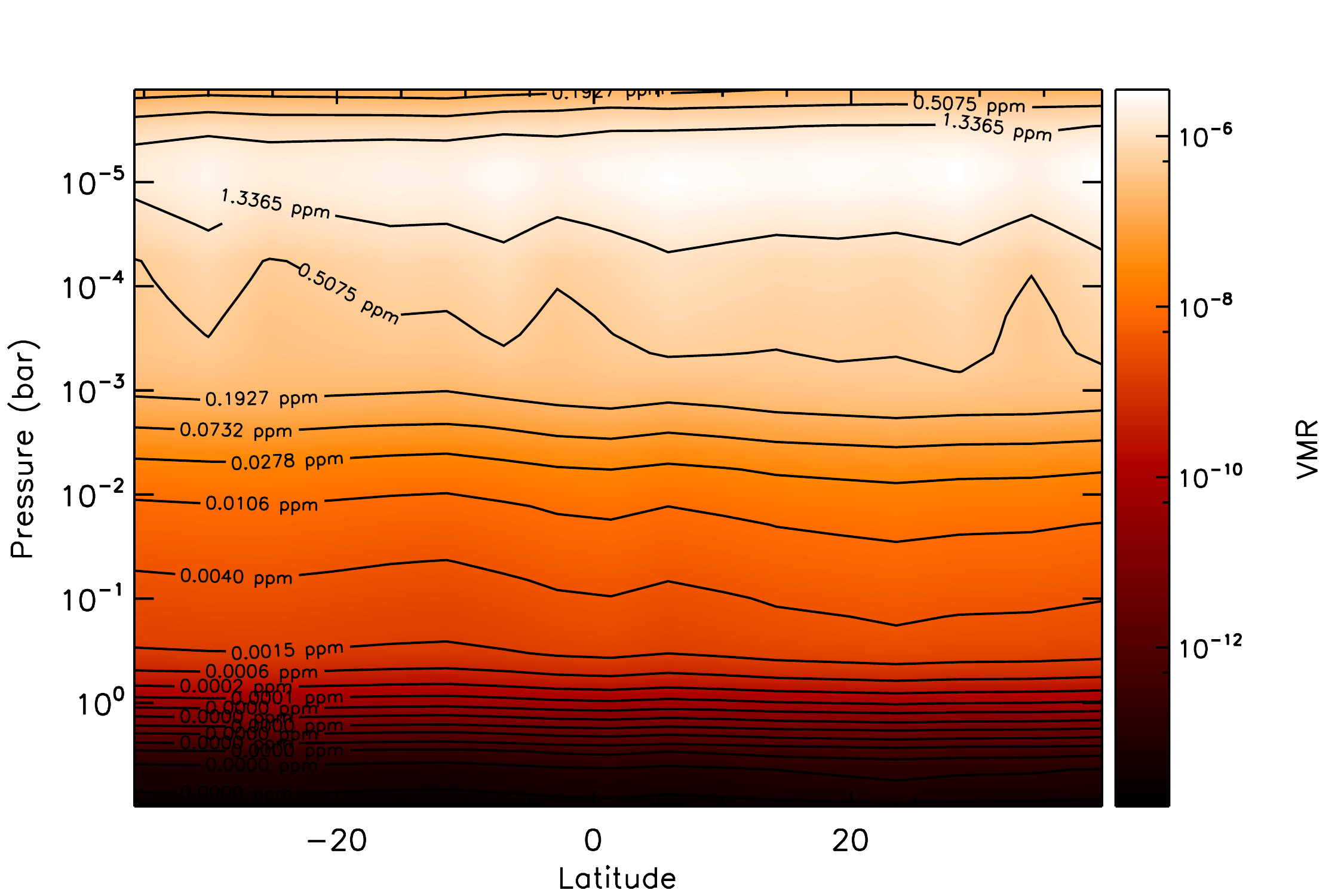}
\caption{ The two dimensional distribution of acetylene compatible with both CIRS and UVIS observations (hypothesis C). At pressures $>$1 mbar the vertical acetylene profile is fixed to values derived from CIRS observations by \cite{2016Icar..278..128F}. \label{ca2dacetylene}}
\end{figure}

\begin{figure}
\centering
\includegraphics[width=0.9\textwidth]{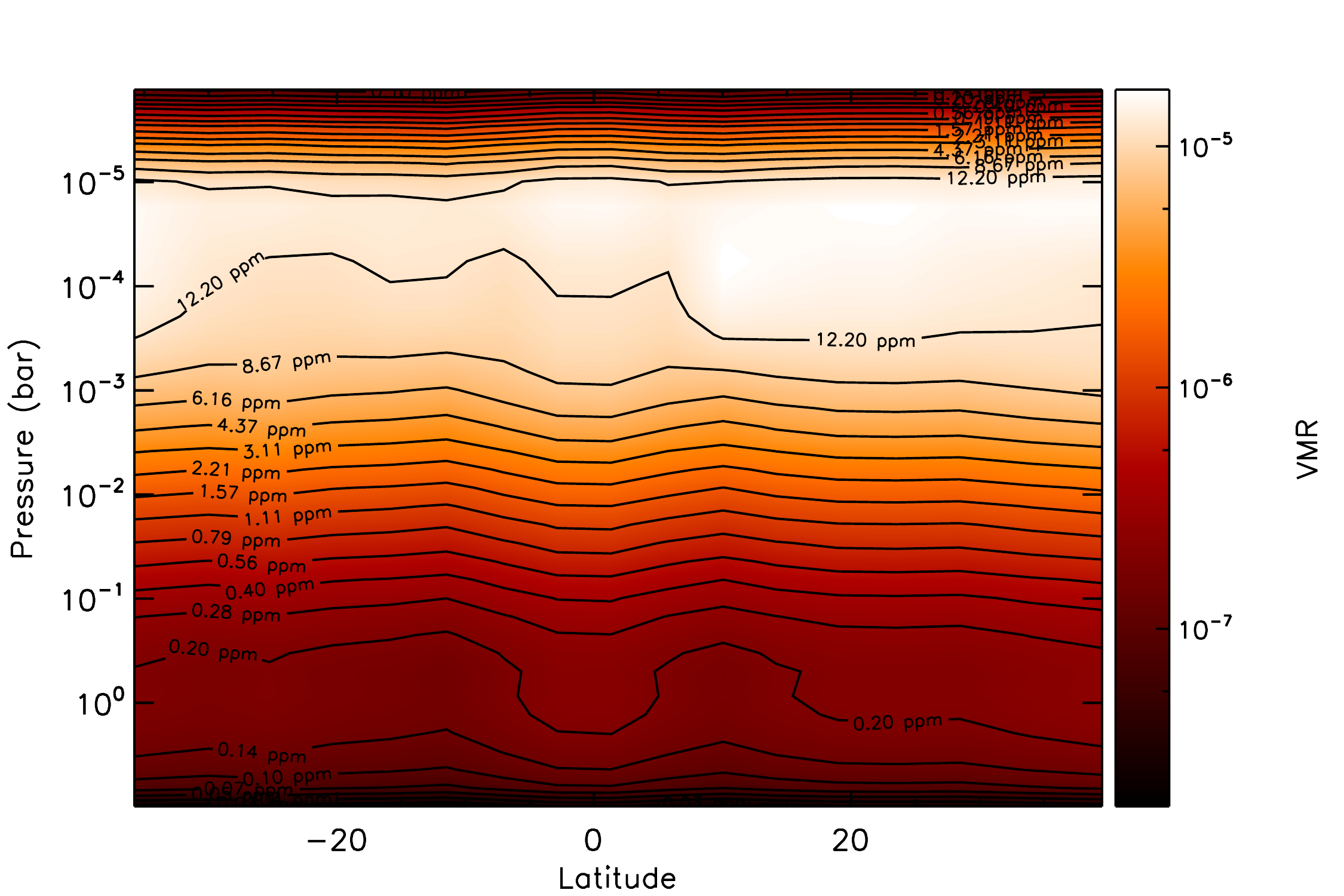}
\caption{The two dimensional distribution of ethane compatible with both CIRS and UVIS observations (hypothesis C). At pressures $>$1 mbar the vertical ethane profile is fixed to values derived from CIRS observations by \cite{2016Icar..278..128F}.  \label{ca2dethane}}
\end{figure}

Figure \ref{ca2dacetylene} and \ref{ca2dethane} show the two dimensional abundance distribution of acetylene and ethane, respectively, as a function of pressure and latitude. These profiles are fixed to the CIRS profiles at pressures $>$1 mbar and are allowed to change at smaller pressure levels, and represent the set of profiles that are the most compatible with {\it both} the ultraviolet UVIS and the mid-infrared CIRS. The factor of two change from 20$^\circ$S to 20$^\circ$N in Figure \ref{cirs_scalings}a can be seen in the asymmetric contours near 10 mbar in Figure \ref{ca2dacetylene}.





\begin{figure}
\centering
\includegraphics[width=0.9\textwidth]{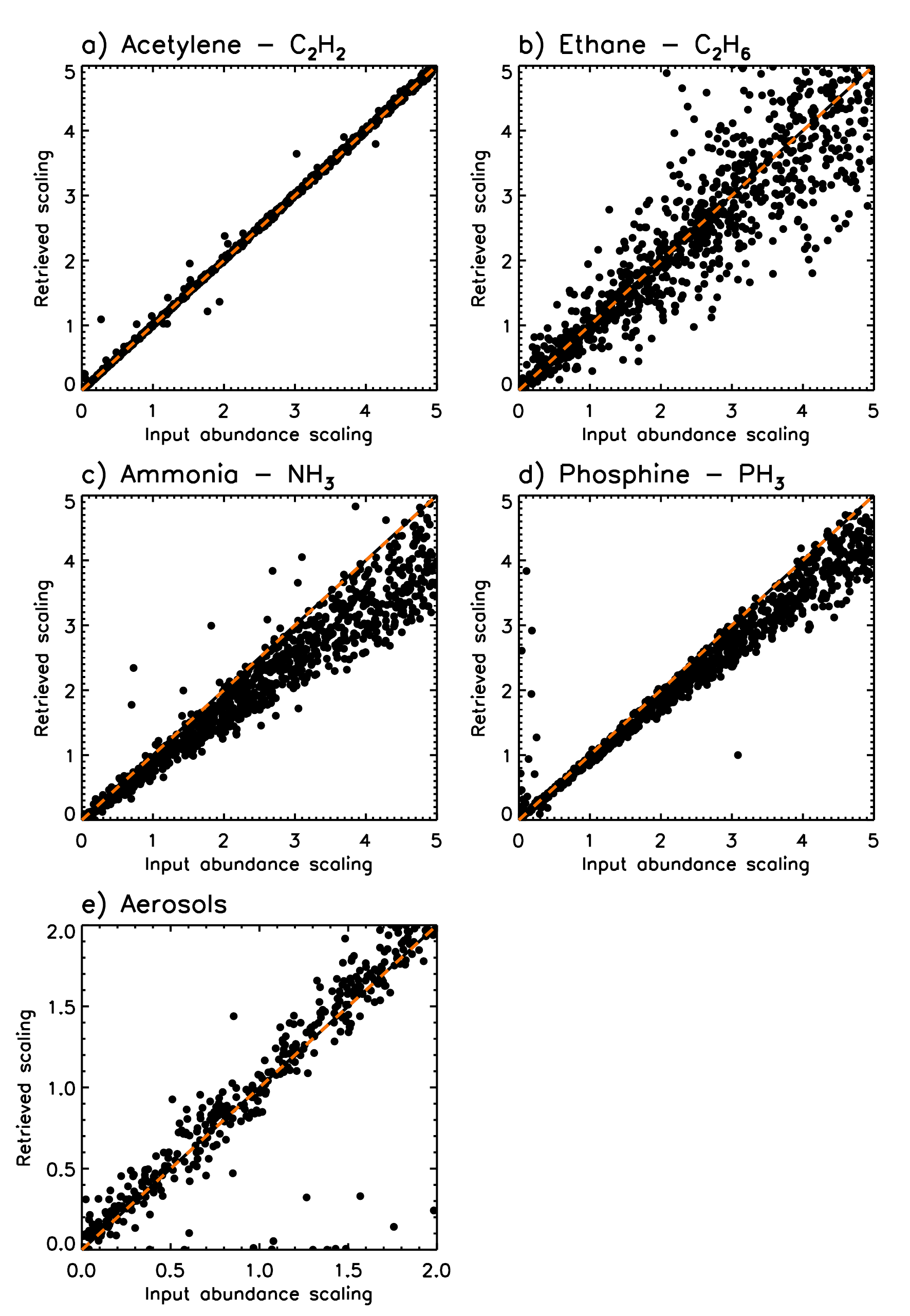}
\caption{A retrieval sensitivity study for the JUICE UVS spectrometer with wavelength coverage up to 0.21 $\mu$m, showing the input abundance scaling of the vertical profiles versus the scaling retrieved using NEMESIS, as described in Section \ref{uvisretsec}. We are able to retrieve all these abundances, to varying degrees of accuracy.  \label{sensuvs}}
\end{figure}

\begin{figure}
\centering
\includegraphics[width=0.9\textwidth]{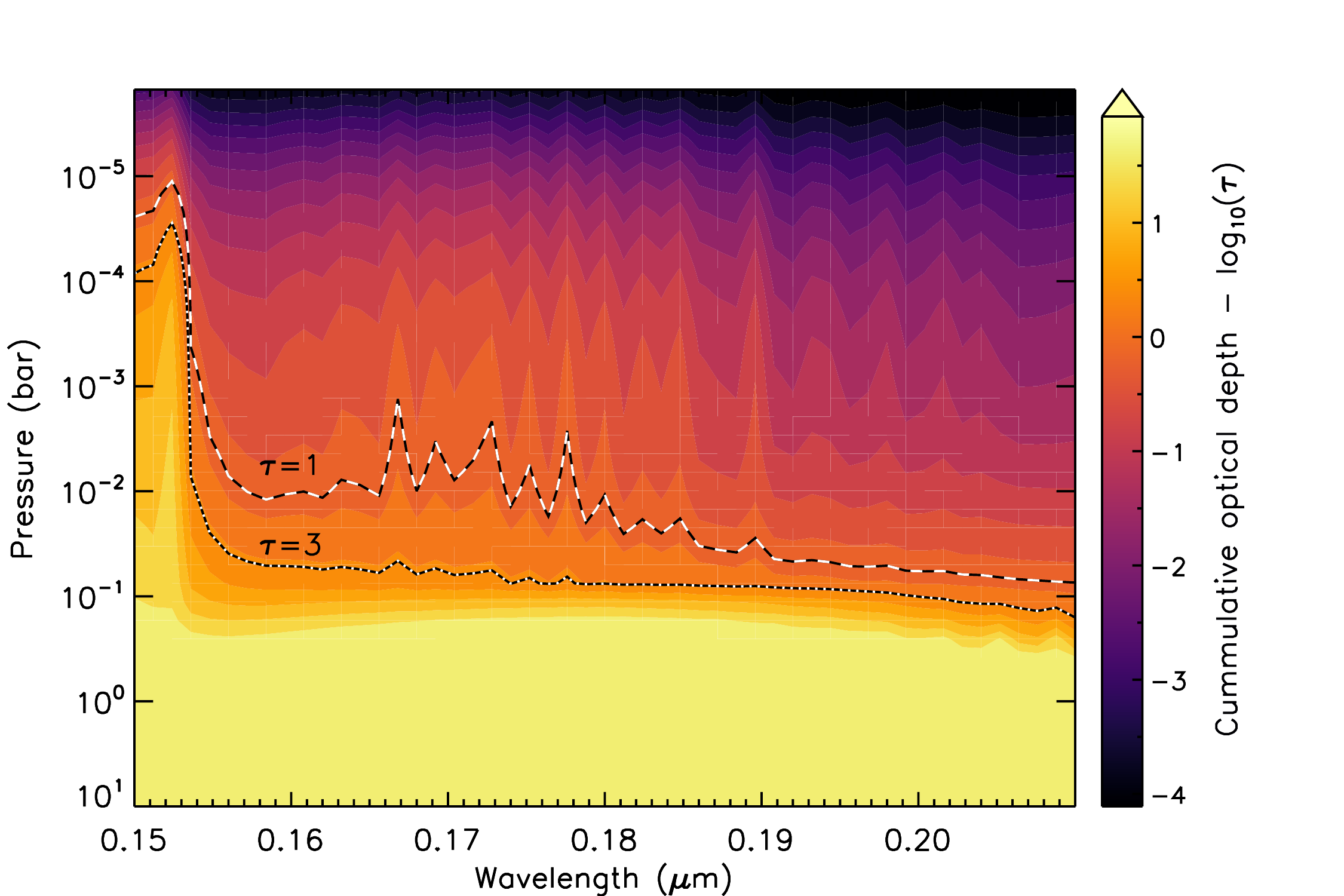}
\caption{The optical depth of the JUICE UVS spectrum as a function of wavelength and pressure, as calculated from the two-way transmission function.  \label{oduvs}}
\end{figure}

\subsection{JUICE UVS}

The Cassini UVIS data analysed in this study were acquired at large distances from Jupiter, translating to relatively low S/N and low spatial resolution. There are, of course, a number of spectrographs that have and will be operating on-board spacecraft bound for the giant planets. This study serves as a proof of concept that ultraviolet reflectance can be analysed by NEMESIS to provide important constraints of abundances in the stratosphere of Jupiter. The European Space Agency's JUICE mission to Jupiter, due to be launched in 2022 and arrive in orbit at the planet in 2029, also carries an ultraviolet spectrograph (UVS). This instrument carries heritage from a number of instruments on both past and present missions, including the New Horizons ALICE \citep{2008SSRv..140..155S} and Juno UVS \citep{2017SSRv..213..447G}. It has a wavelength coverage between 0.05 and 0.21 $\mu$m with a peak spectral resolution of 1.2 nm. Figure \ref{fmadd}a shows that the spectral region between 0.19 and 0.21 is dominated by absorptions by ammonia.

We perform a sensitivity study, as outlined in Section \ref{uvisretsec}, for the spectral resolution and wavelength coverage of JUICE UVS. Figure \ref{oduvs} shows the input abundance scaling versus the scaling retrieved by NEMESIS for acetylene, ethane, ammonia and phoshine, in the same way as Figure \ref{sensuvis}. Similar to Cassini UVIS, JUICE UVS will achieve excellent sensitivity to acetylene and aerosols, with errors of 2\% and 1\% respectively. Ethane can be retrieved moderately well, with a retrieval error of 30\%. The extended wavelength range of UVS allows ammonia and phosphine also be retrieved, both with errors of about 10\%. Note that here the aerosol abundances are only scaled between 0 and 2. This is because the increased aerosol absorption and scattering of aerosols and large abundances of phosphine and ammonia severely reduce the radiance long-ward of 0.19 $\mu$m, darkening the spectrum significantly, making it difficult to perform sensible retrievals. 


The cumulative optical depth for JUICE UVS is shown in Figure \ref{oduvs} -- see Section \ref{odsect} for details on how this is generated. Long-ward of 0.19 $\mu$m unity optical depth occurs between 10 and 100 mbar, providing the sensitivity to retrieve abundances of ammonia and phosphine at the tropopause.

\section{Discussion \label{secdiss}}

Here, we retrieve abundances of $0.78 \pm 0.05$ ppm for acetylene and $15.3 \pm 6.1$ ppm for ethane at 0.1 mbar for Hypothesis C at the NEB during the closest approach UVIS observations of Jupiter. These values are much smaller than any previous abundances derived from ultraviolet reflectance spectra of Jupiter, listed in Table \ref{previous}. Figure \ref{oduvis} shows that the optical depth at the absorption peaks of acetylene reach an optical depth of unity at about 1 mbar, which is the top range of altitudes that UVIS is sensitive to. In fact, Figure \ref{oduvis} shows that the 80-mbar altitude quoted by many the studies in Table \ref{previous} is effectively invisible in the UVIS wavelength range, with the smallest pressure UVIS is able to sample being $\sim$30 mbar at 0.19 $\mu$m. At 5 mbar we retrieve VMR's at the NEB of $56.8 \pm 3.4$ ppb for acetylene and $3.4 \pm 1.4$ ppm for ethane. These values compare reasonably well to the values in Table \ref{previous}. This may suggest that the quoted abundance values in Table \ref{previous} are actually at a higher altitude than previously thought, i.e. closer to 5 mbar than 80-mbar. The use of 80-mbar as a reference pressure for the UV originated in \cite{1974ApJ...187..641T}, but should probably be omitted in future works. Figure \ref{rayonly} shows the optical depth of the Rayleigh scattering process in isolation, broadly consistent with the atmosphere reaching unity optical depth at 80 mbar at 0.2 $\mu$m, decreasing in altitude with increasing wavelength. However, including the absorption of molecular species can increase this altitude significantly, as shown in Figure \ref{oduvis} and Figure \ref{oduvs}. 

The abundance scalings of acetylene in Figure \ref{ca_profiles}a have a symmetric distribution about the equator, whilst the profiles derived from CIRS observations in Figure \ref{cirs_scalings} at lower altitudes contain an enhancement in abundance over the NEB near 20$^{\circ}$N compared to the SEB near 20$^{\circ}$S. Mid-infrared studies showed that this asymmetry was still present in 2013 but the meridional distribution became symmetric in 2017 \citep{2018Icar..305..301M}. The fact that we are not seeing this in the acetylene distribution at pressures $<$1 mbar may indicate that the global circulation is de-coupled between these two altitudes, or that they form part of two different circulation cells.

The Quasi Quadrennial Oscillation \cite[QQO, e.g.][]{1991Natur.354..380L, 2017JGRE..122.2719C} is a semi-regular $\sim$4 year cycle of changing temperatures at the equator, driven largely by gravity waves generated in the turbulent troposphere \citep{2006Icar..180...98S}, which manifests as cold and warm patches propagating downward in altitude, forming an oscillation at a particular pressure level. With Rayleigh scattering being independent of temperature, the ultraviolet reflectance spectrum is only sensitive to temperature in the sense that there is some temperature dependence in the ultraviolet absorption cross-sections and the atmospheric scale height. However, at the cold temperatures observed in the atmosphere of Jupiter, the available cross-section data are effectively single temperature measurements. Therefore, in the absence of future laboratory measurements of cold temperature cross-sections, ultraviolet observations have limited sensitivity to stratospheric temperatures, and are therefore largely blind to any QQO variability in temperature. Both acetylene and ethane show reduced abundances at the equator (Figure \ref{ca_scalings}). This is in contrast to the increased equatorial abundances observed by CIRS (Figure \ref{cirs_scalings}). The vertical profiles of ethane and acetylene derived in this study for Hypothesis C that are consistent with both CIRS and UVIS (e.g. see Figure \ref{ca_profiles}), are consistent with the two instruments sensing different altitudes. Since the CIRS and UVIS abundance scalings at the equator are effectively anti-correlated, this seems to indicate that we are observing QQO signatures in the abundance of both acetylene and ethane. This means that the QQO does not only impact on the temperature field over time, it also has a dynamical effect on the vertical abundances of the hydrocarbons. 

\begin{figure}
\centering
\includegraphics[width=0.9\textwidth]{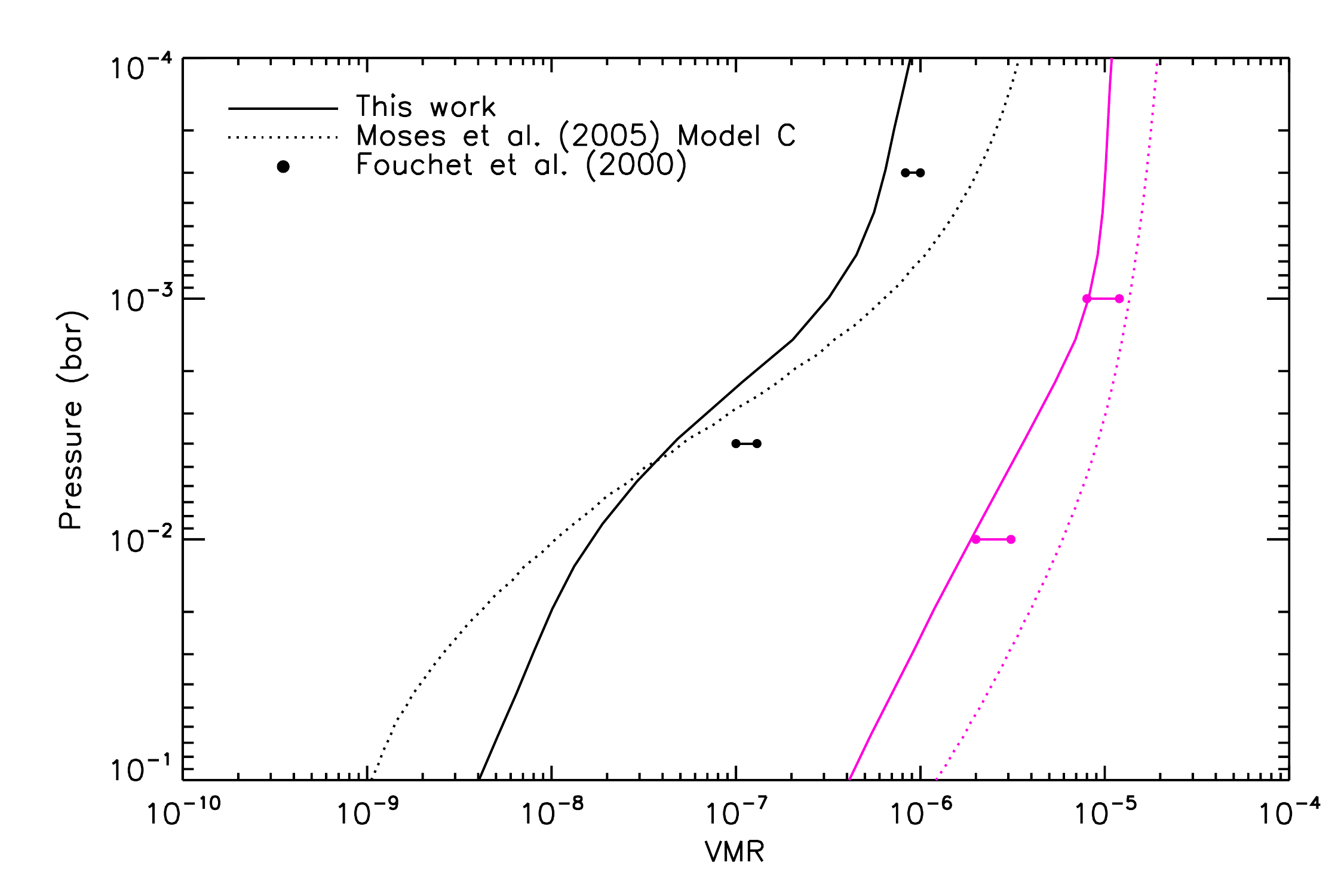}
\caption{Comparison of the retrieved equatorial abundances of acetylene (black) and ethane (pink) of the Cassini UVIS noon retrievals of Hypothesis C with observations in the mbar region by \cite{2000A&A...355L..13F} and \cite{2005JGRE..110.8001M}. Note that the retrievals in this work are held fixed to CIRS values at pressures $>$1 mbar, and the {\it a-priori} profiles used in this study from \cite{2007Icar..188...47N} and \cite{2016Icar..278..128F} have higher abundances of acetylene and ethane at lower altitudes than \cite{2005JGRE..110.8001M}. \label{discomp}}
\end{figure}

The abundances of ethane and acetylene can also be obtained by analysing observations in the mid-infrared. In particular, observations from the Infrared Space Observatory (ISO) operate at a higher spectral resolution than CIRS and sample similar altitudes as considered in this study. The retrieved equatorial abundances of acetylene and ethane of \cite{2000A&A...355L..13F} are shown in Figure \ref{discomp}, along with the vertical abundance retrieved for the Cassini UVIS noon observations, and the model profiles of \cite{2005JGRE..110.8001M}. Similar to the low spatial resolution UVIS noon observations, the ISO observations cover latitudes between $\pm 30^{\circ}$. The ethane abundances (pink) agree well, whilst acetylene (black) abundances retrieved from the UVIS data are lower than those retrieved from ISO data. The ISO data were obtained in May 1997, which suggests that there are significant changes in the abundance of acetylene at pressures $<$1 mbar on the time-scale of years, in agreement with \cite{2018Icar..305..301M}. The photochemical model of \cite{2005JGRE..110.8001M} is plotted as the dotted line in Figure \ref{discomp}, showing agreement within an order of magnitude for both species. The ultraviolet occultation observations of \cite{2010Icar..208..293G} are not directly comparable to the abundances obtained here since they sample pressure levels between 1 and 10 $\mu$bar. 



\section{Conclusions}

We developed the ability of our NEMESIS radiative transfer and retrieval code to model and fit ultraviolet reflectance spectra from Jupiter between 0.15 and 0.19 $\mu$m. In this wavelength region the spectrum is sensitive to acetylene, ethane, and aerosol abundances, and those can be retrieved from the observations - acetylene with $\sim2\%$ confidence, and ethane with $\sim20\%$ confidence. We applied the upgraded model to observations of Jupiter obtained by the Cassini spacecraft during the flyby in late 2000/early 2001. 

The scaled acetylene abundances retrieved from the UVIS observations (hypothesis B) is lower than those from CIRS by a factor of $0.49\pm0.03$ for the noon data and $0.41\pm 0.06$ for the closest approach data. The abundances retrieved for ethane are largely compatible with the CIRS results with a scaling of $\sim$0.95, but are associated with large error bars.


By fixing the abundance profiles of acetylene and ethane at altitudes where CIRS is sensitive ($>$ 1 mbar pressures) but allowing it change at higher altitudes, we are able to produce vertical profiles that are broadly consistent with both the CIRS and UVIS Jupiter flyby data. Thus, combining observations from two different spectral ranges gives us the ability to sample a greater range of pressure levels compared to an individual wavelength range in isolation. With this enhanced vertical sensitivity, the data suggest that the circulation is somehow decoupled above and below the $\sim$1-mbar level, with larger asymmetries in the lower stratosphere than in the upper stratosphere, and the suggestion that the QQO modulates the abundances of hydrocarbons as well as the temperature field.


We also characterised the additional species that we are able to retrieve with the extended wavelength range of the JUICE UVS instrument, adding the ability to ascertain abundances of phosphine and ammonia in the troposphere in addition to stratospheric acetylene and ethane. The JUICE payload lacks a mid-infrared spectrograph akin to Cassini CIRS, so having the modelling tools required to retrieve abundances of acetylene, ethane, ammonia and phosphine from ultraviolet reflectance spectra will be critical for analysing the upper troposphere and stratosphere of Jupiter in the next decade.

\subsection{Ultraviolet ethane absorption cross-sections}

In Section \ref{xsectsect} we show that the ultraviolet cross-section tabulation in \cite{2004JQSRT..85..195C} of the results of \cite{2001ApJ...551L..93L} contains two errors, one at 0.155 $\mu$m and one at 0.157 $\mu$m. These two data-points are too large by a factor of 10, they will result in the retrieved abundance of ethane being too small by a factor of $\sim$10, and the spectrum at 0.155 $\mu$m will appear almost completely absorbed. We urge future authors to treat the tabulation of \cite{2004JQSRT..85..195C} with care, and instead use the corrected values of Table \ref{ethanexsect}.

\section{Acknowledgements}
This work is supported by a European Research Council Consolidator Grant under the European Union’s Horizon 2020 research and innovation program, grant agreement 723890 at the University of Leicester, by a Royal Society Research Fellowship, and by UK Science and Technology Facilities Council (STFC) grant ST/N000749/1. This research used the ALICE High Performance Computing Facility at the University of Leicester. We thank Larry Sromovsky for providing the
code used to generate our Rayleigh-scattering opacities.


\bibliography{refs}{}
\bibliographystyle{aasjournal}



\end{document}